%% file: rotationGRG.tex
\RequirePackage{fix-cm}
\documentclass[smallextended]{svjour3}       
\smartqed  
\usepackage[titletoc,title]{appendix}
\usepackage{url}
\usepackage{graphicx}
 \journalname{GRG}

\newcommand{\citep}{\cite}
\newcommand{\citet}{\cite}

\usepackage[us]{datetime} 

\begin{document}

\title{The rotation problem
}

\titlerunning{The rotation problem
\_\_\_ \hfill \today, \currenttime
}        

\author{R. Michael Jones         
}


\institute{R. Michael Jones \at
              Cooperative Institute for Research in Environmental Sciences \\
              University of Colorado, 
             Boulder, Colorado 80309-0216 \\
             Tel.: +303-492-3845\\
              Fax: +303-492-1149\\
              \email{Michael.Jones@Colorado.edu}             \\
ORCID: 
0000-0002-9493-7456
}

\date{
Draft: \today, \currenttime 
}

\maketitle

\begin{abstract}
\input{jan16A}
\end{abstract}

\input{jan16B}

\begin{acknowledgements}
I thank David Peterson, David Bartlett, and Andrew Hamilton for useful discussion.
\end{acknowledgements}

\appendix
\begin{appendices}
\input{jan16.appendices}

\end{appendices}

\bibliographystyle{spmpsci}      

\input{rotationGRG.bbl}
%
%

\end{document}

%% file: jan16A.tex
Any reasonable form of quantum gravity can explain (by phase interference) why on a large scale, inertial frames seem not to rotate relative to the average matter distribution in the universe without the need for absolute space, finely tuned initial conditions, or without giving up independent degrees of freedom for the gravitational field.  A simple saddlepoint approximation to a path-integral calculation for a perfect fluid cosmology shows that only cosmologies with an average present relative rotation rate smaller than about $T^*H^2 \approx  10^{-71}$  radians per year could contribute significantly to a measurement of relative rotation rate in our universe, where $T^*\approx 10^{-51}$ years is the Planck time and $H \approx 10^{-10}$ yr$^{-1}$ is the present value of the Hubble parameter.  A more detailed calculation (taking into account that with vorticity, flow lines are not normal to surfaces of constant global time, and approximating the action to second order in the mean square vorticity) shows that the saddlepoint at zero vorticity is isolated and that only cosmologies with an average present relative rotation rate smaller than about $T^*H^2 a_1^{1/2} \approx  10^{-73}$  radians per year could contribute significantly to a measurement of relative rotation rate in our universe, where $a_1 \approx 10^{-4}$ is the value of the cosmological scale factor at the time when matter became more significant than radiation in the cosmological expansion.  This is consistent with measurements indicating a present relative rotation rate less than about $10^{-20}$  radians per year.  The observed lack of relative rotation may be evidence for the existence of quantum gravity.

%% file: jan16B.tex
\section{\label{jan16.Intro}Introduction}

Although there are solutions of Einstein's field equations that allow relative rotation of matter and inertial frames, it has long been known that in our universe inertial frames seem not to rotate with respect to the visible stars.  The ``Rotation Problem'' is to explain ``If the universe can rotate, why does it rotate so slowly?''\citep{Ellis-Olive:1983}.  The rotation problem can easily be seen by comparing the rotation of the plane of a Foucault pendulum with the movement of the stars relative to the Earth.  More accurate estimates of the lack of relative rotation are now available, mostly because of the availability of isotropy measurements on the cosmic microwave background radiation.  

For example, Hawking \citet{Hawking69} showed that if the universe contains a large-scale homogeneous vorticity, then the rotation rate corresponding to that vorticity cannot be larger than somewhere between $7 \times 10^{-17}$ rad yr$^{-1}$ and $10^{-14}$ rad yr$^{-1}$ if the universe is closed and about $2\times 10^{-46}/(\mbox{present density in g cm}^{-3})$ if it is open.  Many studies have been done since then resulting in progressively decreasing estimates of the allowed rotation rate.
Table \ref{T2} summarizes some of these measurements.

\begin{table}[h]
\caption{Some measurements of the maximum present relative rotation rate of average inertial frame and average matter distribution.}
\label{T2}
\begin{tabular}{|r|l|}
\hline
\rule[-1 mm]{0 mm}{5 mm}
 max. rotation rate & References  \\
 (radians per year) & \\
\hline
\hline
\rule[-1 mm]{0 mm}{5 mm}
 $2\times 10^{-17}$ (note\footnotemark )&\citet[Hawking, 1969]{Hawking69} \\
 $2\times 10^{-13}$ (note\footnotemark )&\citet[Ellis, 1971, 2009]{Ellis:1971,Ellis:2009} \\
 $1.5\times 10^{-18}$  (note\footnotemark )&\citet[Collins \& Hawking, 1973]{Collins-Hawking:1973b} \\
 $ 10^{-21}$  (note\footnotemark )&\citet[Collins \& Hawking, 1973]{Collins-Hawking:1973b} \\
 $0.7\times 10^{-14}$  (note\footnotemark )&\citet[Barrow et al., 1985]{Barrow-Juszkiewicz-Sonoda:1985} \\
 $4.3\times 10^{-20}$   (note\footnotemark )&\citet[Jaffe et al., 2005, 2006]{Jaffe.et.al:2005,Jaffe.et.al:2006} \\
 $0.7\times 10^{-16}$ (note\footnotemark )&\citet[Ellis, 2006]{Ellis:2006}  \\
 $2\times 10^{-9}$  (note\footnotemark )&\citet[Su \& Chu, 2009]{Su-Chu:2009}  \\
 $3\times 10^{-21}$  (note\footnotemark )&\citet[Saadeh et al., 2016]{PhysRevLett.117.131302}  \\
\hline
\end{tabular}
\end{table}
\addtocounter{footnote}{-8}\footnotetext{for an open universe, assuming about $10^{-29}$ g cm$^{-3}$ for the present density of the universe}
\stepcounter{footnote}\footnotetext{assuming 67.74 km/sec per Mpc for the present value of the Hubble parameter}
\stepcounter{footnote}\footnotetext{if the universe is closed, and if the microwave background was last scattered at a redshift of about 7}
\stepcounter{footnote}\footnotetext{if the universe is closed, and if the microwave background was last scattered at a redshift of 1000}
\stepcounter{footnote}\footnotetext{if the universe is open, assuming 67.74 km/sec per Mpc for the present value of the Hubble parameter}
\stepcounter{footnote}\footnotetext{assuming 67.74 km/sec per Mpc for the present value of the Hubble parameter}
\stepcounter{footnote}\footnotetext{assuming 67.74 km/sec per Mpc for the present value of the Hubble parameter}
\stepcounter{footnote}\footnotetext{maximum rotation rate at the last scattering surface}
\stepcounter{footnote}\footnotetext{assuming 67.74 km/sec per Mpc for the present value of the Hubble parameter}

In General Relativity, gravitation (including inertia) (as expressed by the metric tensor) is determined not only by the distribution of matter (in terms of the stress-energy tensor), but also by initial and boundary conditions.
There are many solutions of Einstein's field equations for General Relativity that have large-scale relative rotation of matter and inertial frames, e.g. \citep{Kramer-Stephani-MacCallum-Herlt:1980,Stephani-Kramer-MacCallum-Hoenselaers-Herlt:2003,EllisMacCallum69,1996gpst.conf..421K,Chechin:2013}.  It is difficult to explain the absence of relative rotation in our universe classically without absolute space (as proposed by Newton) or without assuming very finely tuned initial conditions for the universe.

Ernst Mach \citet{Mach72,Mach72a,Mach33,Mach33a} suggested that inertia might be determined by distant matter.  Various versions of that proposal have come to be known as Mach's principle.  Since we now know (from General Relativity) that inertia is a gravitational force, such an implementation of Mach's principle would require that the gravitational field (or at least part of it) be determined only by its sources (matter) rather than having independent degrees of freedom (in terms of initial and boundary conditions).  

If the many proposals to implement Mach's principle for General Relativity, 
e.g. \citep{Sciama53,SWG69,Gilman70,Raine75,Raine:1981,Raine95}  were correct, then gravitation would behave very differently from the electromagnetic interaction, in that electric and magnetic fields are determined not only from sources (charges and currents), but also from initial and boundary conditions.  

Although General Relativity provides a partial mechanism for Mach's principle through ``frame dragging,'' it is necessary to explain why the initial vorticity is small enough to give the very small present value of vorticity.  A possible explanation for such a small value of initial vorticity may come from some form of quantum gravity.  That is, quantum gravity may provide the mechanism for implementing Mach's principle.

References \citet{Ellis-Olive:1983} and \citet{Braccesi:1988} suggest that inflation might lead to very small shear and rotation rates for our present universe because shear and rotation rate decrease as the universe expands.  Although inflation (and the general expansion) would decrease the relative rotation, it would not be enough to allow for arbitrarily large values of the initial rotation rate.  

Section \ref{jan16.overview} argues why a quantum gravity mechanism to explain the complete absence of vorticity is valid even though we have no generally accepted theory of quantum gravity. 
Section \ref{jan16.path} gives the form of a path integral for a quantum superposition of cosmologies characterized by the rms value $\langle\omega_3\rangle$ of the present vorticity.
Section \ref{jan16.Saddlepoint} shows that there is a saddlepoint at $\langle\omega_3\rangle=0$, and that the only cosmologies contributing a significant amount to the path integral have an upper limit for the rms value of the present vorticity.  
Section \ref{jan16.Evaluation} gives an estimate for the dependence of the Action on $\langle\omega_3\rangle$ and estimates the upper limits on $\langle\omega_3\rangle$.
Section \ref{jan16.summary} makes a saddlepoint approximation to the path integral. 
Section \ref{jan16.discussion} discusses possible problems.  

 Appendix \ref{jan16.amplitude} sets up a calculation of the amplitude for measuring a rotation of the universe in terms of a path-integral calculation.  
 Appendix \ref{jan16.fluid} calculates the action for a perfect fluid.  
Appendix  \ref{jan16.eras} discusses the radiation, matter, and dark-energy eras.
Appendix  \ref{jan16.relative.rotation} estimates the total relative rotation of the universe for the maximum relative rotation rate that contributes significantly to the amplitude for measuring vorticity.
Appendix \ref{jan16.ell} calculates the effect that flow lines are not normal to surfaces of constant global time if vorticity is present.  Specifically, it calculates the scale factor along flow lines as a function of cosmological scale factor to second order in the mean square of the present vorticity, $\langle\omega_3\rangle^2\equiv \overline{\omega_3^2}$.  
Appendix  \ref{jan16.Friedmann} calculates a generalized Friedmann equation to second order in the mean square of the present vorticity when the vorticity is very small.    
A more accurate calculation that takes into account the coupling between vorticity and shear in terms of a vector perturbation \citep[Chapter 10]{Ellis-Maartens-MacCallum:2012} (not shown) would not change the total action significantly.
Appendix  \ref{jan16.main} calculates the approximate action for small vorticity to second order in the mean square of the present vorticity, $\langle\omega_3\rangle^2$. 

We take the speed of light $c$ and Newton's gravitational constant $G$ to be 1 throughout.

\section{\label{jan16.overview}Quantum gravity}

 There are strong reasons why a theory of quantum gravity should exist, e.g.\citep{smolin2014three}, and it is generally believed that such a theory exists.  There are many difficulties with formulating a theory of quantum gravity, some of which are discussed in \citet{isham1975quantum,isham1981quantum}.  Although we do not have a final theory of quantum gravity, and therefore, no universally accepted theory of quantum cosmology,  we have some speculations for theories of quantum gravity, e.g. \citep{DeWitt:1967a,DeWitt:1967b,DeWitt:1967c,Wheeler68,Giulini2009,Kiefer2009,Kiefer2013}.  

However, some calculations (including the present one) can be made without having a full theory of quantum gravity by using a path-integral representation because the action is most likely to dominate over the measure (which we do not know), and the action in the case of vorticity depends only weakly on the exact form of the Lagrangian. 

\section{\label{jan16.path}Path integrals in quantum cosmology}

It is likely that some form of quantum cosmology can explain through phase interference the lack of vorticity. 
One of the standard formulations of quantum cosmology is in terms of path integrals \citep{Hawking79}, in which an initial 3-geometry changes to some final 3-geometry along a ``path'' that is a 4-geometry (i.e., a spacetime, or cosmology).  Thus, each ``path'' is one cosmology.  This is related to a sum-over-histories approach 
\citep{Halliwell:Ortiz:1993,Gell-MannHartle93,Gell-MannHartle90,Hartle92,Hartle1994}.\footnote{also called decoherent histories  
\citep{Halliwell:1988:report,Halliwell:1990,Halliwell:1999,Halliwell:2003,Halliwell:2003a,Halliwell:2005,Halliwell:2005b,Halliwell:2006} or consistent histories  \citep{Griffiths:1984,Griffiths:2002,omnes:1992,omnes:1999}}  

Although in general, the 4-geometries considered in a path integral do not have to be classical cosmologies (that is, solutions of Einstein's field equations), it is known that classical cosmologies usually dominate the path integral, and therefore, here, we shall consider only classical cosmologies in the path integral.

We further restrict cosmologies in our path integral to those cosmologies that differ from the standard cosmological model only in that they have vorticity.  Such cosmologies would have begun with an arbitrary initial vorticity, $\omega_0$, that varied as a function of position, that would then decrease in a known way (Tables \ref{T3} and \ref{T3a})  as the universe expanded during the era when radiation dominated over matter.  
 
 Because radiation density decreases faster than matter density as the universe expands, we eventually reach a time when the density of radiation and matter are equal, and the vorticity has dropped to a value $\omega_1$, which is still a function of position.  From that point on, the vorticity drops at a faster rate (Tables \ref{T3} and \ref{T3a})  as the universe continues to expand until the matter density decreases to a point where the matter term in the Friedmann equation equals the cosmological constant term, and the vorticity has dropped to a value $\omega_2$, which is still a function of position.  
 
 Eventually, we reach the present time, where the vorticity has dropped to a value $\omega_3$, still a function of position.   Because the vorticity depends in a known way on the expansion during the various eras, $\omega_3$ and $\omega_0$ have a known relationship,   that allows us to designate each cosmology by the spatial variation of $\omega_3$.

  \begin{table}[h]
\caption{Radiation, matter, and dark-energy eras \citep[Table 6.1]{Ellis-Maartens-MacCallum:2012} $\ell$ is a scale factor along lines of cosmic flow.   The constant $w=1/3$.  Some of the quantities in this table are a function of position.}
\label{T3}
\begin{tabular}{|c|c|c|c|c|c|c|}
\hline
\rule[-1 mm]{0 mm}{5 mm}
Era & $\ell$ & $\rho \propto$& $\rho =$ & $p$ & $\omega \propto$ & $\omega =$  \\
\hline
\hline
&$\ell_0$&&&&&$\omega_0$\\
Radiation & $\ell_0 \to \ell_1$ & $ \ell^{-4}$& ${3H_3^2}/({8\pi})  (\ell_1 /\ell_2)(\ell_2 /\ell)^{4}$ & $w\rho$ & $\ell^{-1}$ 
& $ \omega_3 ({\ell_3}/{\ell_1}) ({\ell_3}/{\ell})$  \\
&$\ell_1$&&&&&$\omega_1$\\
Matter & $\ell_1 \to \ell_2$ & $ \ell^{-3}$& ${3H_3^2}/({8\pi}) (\ell_2 /\ell)^{3}$ & $0$ & $ \ell^{-2}$ 
& $ \omega_3 ({\ell_3}/{\ell})^2$ \\
&$\ell_2$&&&&&$\omega_2$\\
 Dark energy & $\ell_2 \to \ell_3$ & $ \ell^{-3}$& ${3H_3^2}/({8\pi}) (\ell_2 /\ell)^{3}$ & $0$  & $ \ell^{-2}$ 
 & $ \omega_3 ({\ell_3}/{\ell})^2$ \\
&$\ell_3$&&&&&$\omega_3$\\
\hline
\end{tabular}
\end{table}

  \begin{table}[h]
\caption{Radiation, matter, and dark-energy eras \citep[Table 6.1]{Ellis-Maartens-MacCallum:2012} $a$ is the cosmological scale factor.  $\ell$ is a scale factor along lines of cosmic flow.     In the presence of vorticity, $a$ and $\ell$ differ.  The dimensionless parameter $\delta$ [defined in (\ref{jan16.delta})] is proportional to the square of $\omega_3$.  Some of the quantities in this table are a function of position.}
\label{T3a}
\begin{tabular}{|c|c|c|c|}
\hline
Era & $a$ & $\rho \approx  \left[1+ f_L(a) \delta + f_{LL}(a) \delta^2 \right] \times$  & $\omega \approx  \left[1+ g_\omega(a) \delta + g_{\omega\omega}(a) \delta^2 \right] \times$  \\
\hline
\hline
&$a_0$&&\\
Radiation & $a_0 \to a_1$ &  ${3H_3^2}/({8\pi})  (a_1 /a_2)(a_2 /a)^{4}$  
& $ \omega_3 ({a_3}/{a_1}) ({a_3}/{a})$  \\
&$a_1$&&\\
Matter & $a_1 \to a_2$ &  ${3H_3^2}/({8\pi}) (a_2 /a)^{3}$  
& $ \omega_3 ({a_3}/{a})^2$ \\
&$a_2$&&\\
 Dark energy & $a_2 \to a_3$ &  ${3H_3^2}/({8\pi}) (a_2 /a)^{3}$ 
 & $ \omega_3 ({a_3}/{a})^2$ \\
&$a_3$&&$\omega_3$\\
\hline
\end{tabular}
\end{table}

There is a complex amplitude associated with each ``path'', that is, with each cosmology.  The complex amplitude has a magnitude and a phase, whose values can be calculated from a correct theory of quantum gravity, once that theory is known.  Usually, we expect the value of the phase to dominate the path integral.

Appendix \ref{jan16.amplitude} sets up a calculation of the amplitude for measuring a rotation of the universe in terms of a path-integral calculation.   Appendix  \ref{jan16.main} shows that the action for each cosmology depends only on the present rms vorticity $\langle\omega_3\rangle$, where $\langle\omega_3\rangle^2\equiv\overline{\omega_3^2}$, and the average is a spatial average over the volume within the past light cone.  However, because there are an infinite number of vorticity configurations for each value of the present rms vorticity $\langle\omega_3\rangle$, the path-integral calculation represents an infinite number of integrations.  To correctly deal with that situation, it is useful to make a comparison with a more standard path integral.

To make a path-integral calculation for the propagation of a light wave or a radio wave between a specified source location and a specified observer location in a specified inhomogeneous medium requires an infinite number of parameters to specify the path connecting those two endpoints.  Because of that, a path integral corresponds to an infinite number of integrations.  For most values of the action, there will be an infinite number of paths that have the same action.  However, there are usually only a finite number of discrete paths for which the action is an extremum, and these paths are saddlepoints for the path integral.  This justifies reducing an infinite number of integrations to a small finite number when using saddlepoint approximations to make WKB approximations to path integrals .

The situation in the present case is similar.  For most values of the rms vorticity (and therefore the action, which depends only on the rms vorticity), there are an infinite number of configurations of the spatial variation of vorticity that have the same rms vorticity, and therefore the same action.  However, there is only one configuration at the saddlepoint for which the rms vorticity is zero, and that corresponds to the configuration in which the vorticity is exactly zero everywhere.  Thus, the situation here is exactly analogous to that of the ordinary path integral.  Therefore, it is valid to consider only a simple path integral 
\begin{equation}
\int A(\langle\omega_3\rangle) \exp [i I(\langle\omega_3\rangle)/\hbar]  \mathrm{d} \, \langle\omega_3\rangle
\mbox{ , }
\label{jan16.amplitude.for.process}
\end{equation}
 instead of an infinite number of integrations to represent the amplitude for any process (such as measuring the vorticity),  
 where $A(\langle\omega_3\rangle)$  is a slowly varying function of $\langle\omega_3\rangle$ that gives the magnitude of the contribution of the ``path'' (cosmology), and the phase is equal to the action $I(\langle\omega_3\rangle)$ divided by the reduced Planck's constant $\hbar$.

\section{\label{jan16.Saddlepoint}Saddlepoint}

We expect that $I(\langle\omega_3\rangle)$ should be a smoothly varying function of $\langle\omega_3\rangle$.  Also, we expect that when the rms vorticity $\langle\omega_3\rangle$ is zero, the cosmology will be the standard cosmological model (the standard Robertson-Walker cosmology), and will have the action $I_0$ associated with that model.  Further, because of symmetry\footnote{A cosmology with an average present vorticity equal to $- \omega_3$ about some given axis should be equivalent to a cosmology with an average present vorticity equal to $+\omega_3$ about that same axis, and should therefore have the same action.}, the action $I(\langle\omega_3\rangle)$ should be an even function of $\langle\omega_3\rangle$.  Therefore, the action $I(\langle\omega_3\rangle)$ should look roughly like 
\begin{equation}
I(\langle\omega_3\rangle) \approx 
 I_0 + \hbar \left(\frac{\langle\omega_3\rangle}{\omega_m}\right)^2  f_I(\langle\omega_3\rangle) 
\mbox{ , }
\label{jan16.small.average.omega}
\end{equation}
where $f_I(\langle\omega_3\rangle)$ is some slowly varying dimensionless even function of $\langle\omega_3\rangle$.  Notice that (\ref{jan16.small.average.omega}) reduces to $I_0$, the action for the standard cosmological model, when the rms vorticity $\langle\omega_3\rangle$ is zero, and that (\ref{jan16.small.average.omega}) is an even function of $\langle\omega_3\rangle$ if $f_I(\langle\omega_3\rangle)$ is an even function of $\langle\omega_3\rangle$, as required.  Direct calculation in the appendices shows that (\ref{jan16.small.average.omega}) is correct if $\langle\omega_3\rangle$ is the rms value for $\omega_3$ when taking a spatial average over the past light cone.  

Comparing (\ref{jan16.amplitude.for.process}) and (\ref{jan16.small.average.omega}) shows that (\ref{jan16.amplitude.for.process}) has a saddlepoint at $\langle\omega_3\rangle=0$.  If that saddlepoint is the only significant saddlepoint (and other criteria are satisfied), then the only significant contributions to the path integral (\ref{jan16.amplitude.for.process}) comes from values of the present rms vorticity 
\begin{equation}
\langle\omega_3\rangle=0 \pm \omega_m/\sqrt{f_I(0)}
\mbox{ . }
\label{jan16.limits.omega-3}
\end{equation}
  
Thus, there is no doubt that (\ref{jan16.small.average.omega}) gives the correct behavior for the action for small rms vorticity, nor no doubt that (\ref{jan16.limits.omega-3}) gives the limits for the present rms vorticity of cosmologies that contribute significantly to the path integral in (\ref{jan16.amplitude.for.process}).  The only question is in the values for $\omega_m$ and for the function $f_I(\langle\omega_3\rangle)$.

\section{\label{jan16.Evaluation}Evaluation of the action}

Estimating the value of $\omega_m$ is straightforward.  Estimating the function $f_I(\langle\omega_3\rangle)$ is difficult, as is the task of estimating the dependence of the action on the rms vorticity for all values of the rms vorticity.  The appendices make such estimates.   

If there is relative rotation of the  matter distribution and the  inertial frame, then  density and pressure will depend on the scale length $\ell$ along flow lines rather than depend on the cosmological scale factor $a$.  The significance is that the cosmological scale factor $a$ is a function of global time, but scale length $\ell$ along flow lines is not normal to surfaces of constant global time.  This causes the calculation to be much more complicated.

Appendix \ref{jan16.fluid} calculates the action in terms of a 4-volume integral of the Lagrangian for a perfect fluid plus a surface term.  For some cosmological models, the surface term can be absorbed into the 4-volume integral of an effective Lagrangian.  It is argued that a general effective Lagrangian for solutions of Einstein's field equations can be expressed as a linear combination of pressure, density, and the cosmological constant with constant coefficients.  Specifically, equation (\ref{jan16.L.tilde}) in appendix \ref{jan16.fluid}  gives an effective Lagrangian of the form \begin{equation} \tilde{L} = \alpha_1 p + \alpha_2 \rho + \alpha_3 \Lambda \mbox{ , } \label{jan16.L.tilde.copy} \end{equation} where $p$ is pressure, $\rho$ is density, $\Lambda$ is the cosmological constant, and $\alpha_1$, $\alpha_2$, and $\alpha_3$ are dimensionless constants of order unity.  

Appendix  \ref{jan16.eras} discusses the radiation, matter, and dark-energy eras because the effective Lagrangian would depend differently on the cosmological scale factor in the three eras.  To a first approximation, we can neglect all but the dominant term in each of the three eras.  

$\omega_m$ is the dominant factor that gives the effect of vorticity on the action in (\ref{jan16.small.average.omega}).  The main effect of vorticity on the action can be found by a straightforward calculation of the action integral.  The time integral to give the action is converted to an integral over the cosmological scale factor using a generalization of the Friedmann equation that includes vorticity.  The spatial integral to give the action is approximated by averaging the integrand times the spatial volume within the past light cone, which is proportional to the cube of present radius of the visible universe.  The calculation is expanded to first order in the mean square vorticity.  

The result is given by (\ref{jan16.omegaSub.m}) in appendix \ref{jan16.main} as
\begin{equation}
 \omega_m =  \left( \frac{ \hbar H_3 }{ r_3^3 } \right)^{1/2}
=  \frac{T^*}{r_3}\sqrt{\frac{H_3}{r_3}}
\approx  T^* H_3^2
 \approx    10^{-71} \mbox{rad yr$^{-1}$}
\mbox{ , }
\label{jan16.average.omegaSub.m.2}
\end{equation}
where $H_3\approx \sqrt{\Lambda/3}\approx 10^{-10}$ yr$^{-1}$ is the present value of the Hubble parameter, $r_3$ is the present radius of the visible universe (which we approximate by the inverse of the Hubble parameter), $T^*\approx 10^{-51}$ years is the Planck time, and $\Lambda$ is the cosmological constant. 

Including vorticity in the calculation of relative rotation (even approximately) is tedious, even though straightforward.  
Appendix \ref{jan16.main} calculates the total action to second order in the mean square of the present vorticity.  

The result is given by  (\ref{jan16.6.1.c.sec.order.2}) in appendix  \ref{jan16.main}, which shows that to lowest order in $\langle\omega_3\rangle/H_3$
\begin{eqnarray}
&
f_I(\langle\omega_3\rangle) \approx 
\left(\frac{a_3}{a_1}\right)
\left[C_I + \frac{\langle\omega_3\rangle^2+\sigma_\omega^2/\langle\omega_3\rangle^2}{H_3^2} 
\left(\frac{a_3}{a_1}\right)
C_{II}\right] 
\nonumber \\ 
&
\approx 
\left(\frac{a_3}{a_1}\right)
C_I 
\mbox{ for $\langle\omega_3\rangle \ll 10^{59} \omega_m$, }
\label{jan16.f2.new}
\end{eqnarray}
where $\sigma_\omega^2$ is the variance of $\omega_3^2$, $a_1$ is the value of the cosmological scale factor when the density of radiation was equal to the density of matter, $a_3=1$ is the present value of the cosmological scale factor, and $C_I$ and $C_{II}$ are dimensionless constants of order unity\footnote{within a factor of a hundred, or so.  Compared to a factor of $10^{59}$, a hundred is of order unity.}.  Thus, $f_I(\langle\omega_3\rangle)$ is essentially constant in a large region surrounding the saddlepoint at $\langle\omega_3\rangle=0$.  \footnote{ A simple calculation that neglects the fact that flow lines are not normal to surfaces of constant global time \citep{Jones:rotation:problem:2017v1} shows that only cosmologies whose present relative rotation rate is less than about $  \omega_m \approx T^* H_3^2 \approx 10^{-71}$ radians per year would contribute significantly to a path integral for a measurement of relative rotation rate.  
That is, the simple calculation gave $f_I(\langle\omega_3\rangle) \approx 1$.  That the resulting maximum total rotation since the initial singularity would be less than about  $10^{-61}$ radians suggests that neglecting that flow lines are not normal to surfaces of constant global time should be a good approximation.}

\section{\label{jan16.summary}Saddlepoint approximation}

The saddlepoint at $\langle\omega_3\rangle=0$ in  (\ref{jan16.amplitude.for.process}) is isolated from other saddlepoints and any possible non-analytic points as shown by (\ref{jan16.f2.new}).  The integral in (\ref{jan16.amplitude.for.process}) can be approximated by a saddlepoint integration to give

\begin{eqnarray}
&
A(0)
{\omega_m \sqrt{\pi}}/{\sqrt{f_I(0)}} e^{i\pi/4}
\nonumber \\ 
&
\mbox{ for $|\langle\omega_3\rangle|< \frac{\omega_m}{\sqrt{|f_I(0)|}} \approx \frac{T^* H_3^2}{\sqrt{|f_I(0)|}}
\approx \frac{T^* H_3^2}{ \sqrt{|C_I|}}\left(\frac{a_1}{a_3}\right)^{1/2}
\approx \frac{10^{-73}}{ \sqrt{|C_I|}}
$ 
rad/year, 
}
\label{jan16.integration4b.2dup.2b}
\end{eqnarray}
$ \approx 0$ otherwise.  

Table \ref{T1} gives possible approximate values for $C_I$ and $C_{II}$, depending on reasonable values for $\alpha$, $\alpha_1$, $\alpha_2$,  $\alpha_3$, and $\alpha_4$\footnote{where $\alpha_4$ is an arbitrary constant of integration in (\ref{jan16.H.omega.ell}) in appendix \ref{jan16.Friedmann}, whose value is most likely zero.}, and depending on whether the surface term is included in the effective Lagrangian.  That $C_I$ can be either positive or negative is not a problem, because the path of integration along the real $\langle\omega_3\rangle$ axis is a stationary-phase path.  That the most likely value for $\alpha_4$ is zero gives $|C_I|\approx 1$, which gives from (\ref{jan16.integration4b.2dup.2b}), $|\langle\omega_3\rangle|< \approx {10^{-73}}$ rad/year.  However, larger values of $\alpha_4$ from Table \ref{T1} could give larger values for $|C_I|$, which could give limits on $|\langle\omega_3\rangle|$ as small as about ${10^{-75}}$ rad/year. 
  
  \begin{table}[h]
\caption{Possible approximate values of $C_I$ from (\ref{jan16.6.f.I.7e}) and $C_{II}$ from (\ref{jan16.6.f.II.5}) in appendix \ref{jan16.main}.  The parameter $\alpha$ is defined in (\ref{jan16.6.7a}).  The parameters $\alpha_1$, $\alpha_2$, and $\alpha_3$ are defined in (\ref{jan16.L.tilde.copy}).  The constant of integration $\alpha_4$ is defined in (\ref{jan16.H.f1}), and its most probable value is zero.}
\label{T1}
\begin{tabular}{|c|c|c|c|c|c|c|c|}
\hline
\rule[-1 mm]{0 mm}{5 mm}
References & $\alpha$ & $\alpha_1$ & $\alpha_2$ & $\alpha_3$ &$\alpha_4$ & $C_I$ & $C_{II}$ \\
\hline
\hline
\rule[-1 mm]{0 mm}{5 mm}
\cite{SchutzSorkin77}  & $0$ & $-
{3}/{2}$ & $
{3}/{2}$ & $
{1}/{8\pi}$  && $1 - 686 \alpha_4$ & $-453-8927 \alpha_4+1.04\times10^{6}\alpha_4^2$ \\
  & $0$ & $- {3}/{2}$ & $ {3}/{2}$ & $ {1}/{8\pi}$  & $0$ & $1 $& $-453 $ \\
  & $0$ & $- {3}/{2}$ & $ {3}/{2}$ & $ {1}/{8\pi}$  & $10^{-1} $ & $-68 $ & $9054 $ \\
  & $0$ & $- {3}/{2}$ & $ {3}/{2}$ & $ {1}/{8\pi}$  & $1$ & $- 685 $& $1.03\times 10^{6} $ \\
including & $0$ & $0$ & $0$ & $- {1}/{4\pi}$  && $ -2 + 471
 \alpha_4$ &$-9+204\alpha_4-2.08\times10^{6}\alpha_4^2 $\\
the  & $0$ & $0$ & $0$ & $- {1}/{4\pi}$  & $0$ & $-2  $&-9 \\
surface & $0$ & $0$ & $0$ & $- {1}/{4\pi}$  & $10^{-1} $ & $ 45 $&$-2.08\times 10^{4}$ \\
term & $0$ & $0$ & $0$ & $- {1}/{4\pi}$  & $1$ & $ 469 $&$-2.08\times 10^{6}$ \\
 \hline
\rule[-1 mm]{0 mm}{5 mm}
\cite{MacCallum-Taub:1972}, \cite{Schutz76} & $1$ & $- {1}/{2}$ & $ {1}/{2}$ & $ {1}/{8\pi}$ & & 
$ 1 - 384 \alpha_4$&  $-148-3044  \alpha_4+1.05\times10^{6}\alpha_4^2$\\
 & $1$ & $- {1}/{2}$ & $ {1}/{2}$ & $ {1}/{8\pi}$ & $0$ & $ 1 $ &-148\\
 & $1$ & $- {1}/{2}$ & $ {1}/{2}$ & $ {1}/{8\pi}$ & $10^{-1}$ & $ -37$ &$ 10^{4}$ \\
 & $1$ & $- {1}/{2}$ & $ {1}/{2}$ & $ {1}/{8\pi}$ & $1$ & $ - 383$ &$1.05\times 10^{6}$\\
including & $1$ & $1$ & $-1$ & $- {1}/{4\pi}$ & & $-2 + 769 \alpha_4$ &$296+6087 \alpha_4-2.06\times10^{6}\alpha_4^2$\\
the & $1$ & $1$ & $-1$ & $- {1}/{4\pi}$ & $0$ & $-2 $ &296\\
 surface & $1$ & $1$ & $-1$ & $- {1}/{4\pi}$ & $10^{-1}$ & $75 $ &$-1.97\times10^{4}$\\
 term & $1$ & $1$ & $-1$ & $- {1}/{4\pi}$ & $1$ & $ 767 $ &$-2.05\times10^{6}$\\
\hline
\end{tabular}
\end{table}

\section{\label{jan16.discussion}Discussion}

The calculation that the only cosmologies that contribute significantly to a measurement of rms vorticity have a present rms vorticity less than about $10^{-73}$ to $10^{-71}$ radians per year is fairly robust.  Although reasonable different choices for the Lagrangian might possibly change that result by a factor of a hundred or so, the result is so much smaller than the upper limit of $10^{-20}$ radians per year set by measurements, that there seems no doubt that we have found the correct reason for the lack of any significant measurement of rms vorticity.

That flow lines are not in general normal to surfaces of constant global time in the presence of vorticity makes the calculation more complicated.  Detailed calculation in which the action is calculated through second order in the square of vorticity shows that the saddlepoint at zero vorticity is isolated from any other saddlepoints and from any non-analytic points.  

A correct calculation of the action should include the coupling between vorticity and shear in terms of vector perturbations \citep[Chapter 10]{Ellis-Maartens-MacCallum:2012} \citep[Chapter 29]{Hamilton:2019}.  Although neglecting the coupling between vorticity and shear could be justified simply because the coupling would be of second order, a more accurate calculation that takes into account the coupling between vorticity and shear in terms of a vector perturbation \citep[Chapter 10]{Ellis-Maartens-MacCallum:2012} (not shown) would not change the total action significantly.  

The observed lack of relative rotation may be evidence for the existence of quantum gravity.

%% file: jan16.appendices.tex
\section{\label{jan16.amplitude}Amplitude for measuring a rotation of the universe}

The amplitude for measuring a particular value for some quantity is equal to the amplitude for measuring that value given a particular 4-geometry times the amplitude for that 4-geometry, and then we sum over all 4-geometries.

For example, following \citet{Hartle-Hawking83}, the amplitude for the 3-geometry and matter field to be fixed at specified values on two spacelike hypersurfaces is
\begin{equation}
\langle^{(3)}\mathcal{G}_f,\phi_f|^{(3)}\mathcal{G}_i,\phi_i\rangle
= \int
\psi[^{(4)}\mathcal{G},\phi]
\, \mathcal{D} {^{(4)}\mathcal{G}} \, \mathcal{D} \phi \, 
\mbox{ , }
\label{jan16.relative-rotation0}
\end{equation}
where the integral is over all 4-geometries and field configurations that match the given values on the two spacelike hypersurfaces, and 
\begin{equation}
\psi[^{(4)}\mathcal{G},\phi] \equiv \exp{(i I[^{(4)}\mathcal{G},\phi]/\hbar)}
\mbox{  }
\label{jan16.psi.relative-rotation0}
\end{equation}
is the contribution of the 4-geometry $^{(4)}\mathcal{G}$ and matter field $\phi$ on that 4-geometry to the path integral, where $I[^{(4)}\mathcal{G},\phi]$ is the action.  The proper time between the two hypersurfaces is not specified.  A correct theory of quantum gravity would be necessary to specify the measures $\mathcal{D} {^{(4)}\mathcal{G}}$ and $\mathcal{D} \phi$, but that will not be necessary for the purposes here.  Hartle and Hawking \citet{Hartle-Hawking83} restricted the integration in (\ref{jan16.relative-rotation0}) to compact (closed) 4-geometries, but (\ref{jan16.relative-rotation0}) can be applied to open 4-geometries if that is done carefully.

 Equation (\ref{jan16.relative-rotation0}) is a path integral.  In this case, the ``path'' is the sequence of 3-geometries that form the 4-geometry ${^{(4)}\mathcal{G}}$.  Thus, each 4-geometry is one ``path.''  The space in which these paths exist is often referred to as superspace, e.g. \citep{MTW:1973}.  
 As pointed out by Hajicek \citet{Hajicek:1986}, there are two kinds of path integrals: those in which the time is specified at the endpoints, and those in which the time is not specified.  The path integral in (\ref{jan16.relative-rotation0}) is the latter.  References \citet{Hajicek:1986} and \citet{Kiefer:1991} consider refinements to the path integral in (\ref{jan16.relative-rotation0}), but such refinements are not necessary here.
 
 Because of diffeomorphisms, a given 4-geometry can be specified by different metrics that are connected by coordinate transformations.  This makes it difficult to avoid duplications when making path integral calculations.  We avoid that difficulty here by considering only simple models.
 
Let $\psi_i(^{(3)}\mathcal{G}_i, \phi_i)$ be the amplitude that the 3-geometry was $^{(3)}\mathcal{G}_i$ on some initial space-like hypersurface and that the matter fields on that 3-geometry were $\phi_i$.  Let $\psi_f(^{(3)}\mathcal{G}_f, \phi_f)$ be the amplitude that the 3-geometry is $^{(3)}\mathcal{G}_f$ on some final space-like hypersurface and that the matter fields on that 3-geometry are $\phi_f$.  
 Then, we have
 \begin{equation}
\psi_f(^{(3)}\mathcal{G}_f, \phi_f) 
= \int
\langle^{(3)}\mathcal{G}_f,\phi_f|^{(3)}\mathcal{G}_i,\phi_i\rangle
\psi_i(^{(3)}\mathcal{G}_i, \phi_i) 
\, \mathcal{D} {^{(3)}\mathcal{G}_i} \, \mathcal{D} \phi_i \, 
\mbox{ . }
\label{jan16.relative-rotation0d}
\end{equation}
The condition that there are not finely tuned initial conditions is equivalent to $\psi_i(^{(3)}\mathcal{G}_i, \phi_i)$ being a broad wave function.  

Substituting (\ref{jan16.relative-rotation0}) into (\ref{jan16.relative-rotation0d}) gives
 \begin{equation}
\psi_f(^{(3)}\mathcal{G}_f, \phi_f) 
= \int
\int
\psi[^{(4)}\mathcal{G},\phi]
\, \mathcal{D} {^{(4)}\mathcal{G}} \, \mathcal{D} \phi \, 
\psi_i(^{(3)}\mathcal{G}_i, \phi_i) 
\, \mathcal{D} {^{(3)}\mathcal{G}_i} \, \mathcal{D} \phi_i \, 
\mbox{ . }
\label{jan16.relative-rotation0e}
\end{equation}

Although in (\ref{jan16.relative-rotation0e}), the integration is over all possible 4-geometries, not just classical 4-geometries, 
the main contribution to the integral (in most cases) comes from classical 4-geometries, e.g. 
\citep{Halliwell-Hartle:1990,Kiefer:1991}.  Thus, we shall now restrict (\ref{jan16.relative-rotation0e}) to be an integration over classical 4-geometries.  This is appropriate for our purposes, in any case, since we are trying to explain why we do not measure relative rotation of matter and inertial frames in what appears to be a classical universe.

In principle, the idea is very simple.  Any measurement to determine the inertial frame will give a result that depends on the 4-geometry.  If several 4-geometries contribute significantly to an amplitude, such as in (\ref{jan16.relative-rotation0e}), then any measurement to determine an inertial frame might give the inertial frame corresponding to any one of those 4-geometries.  However, the probability for the result being a particular inertial frame will depend on the contribution of the corresponding 4-geometry to calculations such as that in (\ref{jan16.relative-rotation0e}).

In this calculation, we consider 4-geometries characterized by a parameter ${\langle\omega_3\rangle}$ which we take to be the rms vorticity on the final space-like hypersurface.  Thus, we can rewrite (\ref{jan16.relative-rotation0e}) for our purposes as
 \begin{equation}
\psi_f(^{(3)}\mathcal{G}_f, \phi_f) 
= 
\int
 \int_{-\infty}^{\infty} 
\psi_i(^{(3)}\mathcal{G}_i, \phi_i) 
\psi[^{(4)}\mathcal{G},\phi;\langle\omega_3\rangle] 
\, \mathrm{d}{\langle\omega_3\rangle} 
\, \mathcal{D} {^{(3)}\mathcal{G}_i} \, \mathcal{D} \phi_i \, 
\mbox{ . }
\label{jan16.relative-rotation0c}
\end{equation}

The integral in (\ref{jan16.relative-rotation0c}) is still a path integral.  In this case, each value of $\langle\omega_3\rangle$ specifies one ``path'', in that it specifies one 4-geometry, and that specifies one sequence of 3-geometries.  The space of ``paths'' in this case is often referred to as a mini-superspace because it is restricted to a much smaller space of 4-geometries.  The parameter ${\langle\omega_3\rangle}$, classically determined by initial conditions on the 4-geometry, represents an independent degree-of-freedom of the gravitational field.  

Actually, taking the ${\langle\omega_3\rangle}$ integration from $-\infty$ to $\infty$ in (\ref{jan16.relative-rotation0c}) is not physically realistic, and might lead to problems if the infinite endpoints contribute significantly to the integration.  The largest relative rotation that could possibly be considered without having a theory of quantum gravity would be one rotation of the universe in a Planck time.  This would correspond to taking the maximum value of ${\langle\omega_3\rangle}$ to be the reciprocal of the Planck time, $T^*$, or ${\omega}_{\mbox{max}} \approx 10^{44} \mbox{sec}^{-1}$.  Thus, we can rewrite (\ref{jan16.relative-rotation0c}) as
 \begin{equation}
\psi_f(^{(3)}\mathcal{G}_f, \phi_f)
\approx \int \int_{-{\omega}_{\mbox{max}}}^{{\omega}_{\mbox{max}}} 
\psi_i(^{(3)}\mathcal{G}_i, \phi_i) 
\psi[^{(4)}\mathcal{G},\phi;\langle\omega_3\rangle] \, \mathrm{d}{\langle\omega_3\rangle} \, \mathcal{D} {^{(3)}\mathcal{G}_i} \, \mathcal{D} \phi_i \, 
\mbox{ . }
\label{jan16.relative-rotation}
\end{equation}
  
We anticipate that the properties of $\psi[^{(4)}\mathcal{G},\phi;\langle\omega_3\rangle]$ will dominate the integral in (\ref{jan16.relative-rotation}), so we shall start with 
\begin{equation}
 \psi[^{(4)}\mathcal{G},\phi;\langle\omega_3\rangle] \approx e^{i I(\langle\omega_3\rangle)/\hbar}
\mbox{ , }
\label{jan16.Psi}
\end{equation}
where $I(\langle\omega_3\rangle)$ is the action.  

Either a stationary-phase path or a steepest-descent path could be used when making the saddlepoint approximation 
\citep{Budden:1961:book,Felsen-Marcuvitz:1973:book,Halliwell-Louko:1989}, but here, we use a stationary-phase path.  Halliwell \citet{Halliwell:1988} gives an example of a more detailed path-integral calculation of quantum gravity.

\section{\label{jan16.fluid}Action for a perfect fluid}

	We can take the action in (\ref{jan16.Psi}) to be
\begin{equation}
I = \int (-g^{(4)})^{1/2} 
L
 \mathrm{d}^4x 
+ \frac{1}{8\pi} \int (g^{(3)})^{1/2} K \mathrm{d}^3x
\mbox{ , }
\label{jan16.6.1}
\end{equation}
where 
\begin{equation}
L= L_{\mbox{geom}} + L_{\mbox{matter}}
\mbox{  }
\label{jan16.Lagrangian}
\end{equation}
is the Lagrangian, and the surface term is necessary to insure consistency if the action integral is broken into parts \cite{York72,Hawking79}.   The quantity 
\begin{equation}
K 
= g^{(3)ij} K_{ij} 
= - \frac{1}{2} g^{(3)ij} 
 \frac{\partial g_{ij}^{(3)}}{\partial t}
\label{jan16.6.2}
\end{equation}
is the trace of the extrinsic curvature, 
where $g_{ij}^{(3)}$ is the 3-metric.  In this example, we take the Lagrangian for the geometry as
\begin{equation}
L_{\mbox{geom}} = \frac{R^{(4)}-2\Lambda}{16\pi}
\mbox{ , }
\label{jan16.6.4}
\end{equation}
where $R^{(4)}$ is the four-dimensional scalar curvature and $\Lambda$ is the cosmological constant.  

	For a perfect fluid, the energy momentum tensor is
\begin{equation}
T^{\mu\nu} = (\rho + p) u^{\mu} u^{\nu} + p g^{\mu\nu}
\mbox{ , }
\label{jan16.6.5}
\end{equation}
where p is the pressure, $\rho$ is the density, and $u$ is the
4-velocity.  For solutions to Einstein's field equations\footnote{using the convention of \cite{MTW:1973}.  The field equations (\ref{jan16.Field.eqs}) take on a slightly different interpretation according to \cite{RYSKIN:2015:258}.  However, the development here is valid for either interpretation.} 
\begin{equation}
R^{\mu\nu} = 8\pi (T^{\mu\nu} - \frac{1}{2}  g^{\mu\nu} T) + \Lambda g^{\mu\nu}
\mbox{  }
\label{jan16.Field.eqs}
\end{equation}
for a perfect fluid, (\ref{jan16.6.4}) becomes
\begin{equation}
L_{\mbox{geom}} = \frac{1}{2} \rho - \frac{3}{2} p +
\frac{\Lambda}{8\pi}
\mbox{ , }
\label{jan16.6.6}
\end{equation}
and we can take the Lagrangian for the matter as 
\begin{equation}
L_{\mbox{matter}} = \rho + \alpha(p-\rho)
\mbox{ , }
\label{jan16.6.7a} 
\end{equation}
where $\alpha$ is a constant, and we can take $\alpha = 0$ \citep{SchutzSorkin77}, $\alpha = 1$ \citep{MacCallum-Taub:1972,Schutz76}, or $\alpha = \frac{3}{2}$ (from combining (\ref{jan16.6.5}) with \cite[eq. 21.33a]{MTW:1973}).  However, as we shall see,  the final result is insensitive to the exact form of the Lagrangian.  Substituting (\ref{jan16.6.6}) and (\ref{jan16.6.7a}) into (\ref{jan16.Lagrangian}) gives 
\begin{equation}
{L} = (\alpha - \frac{3}{2})(p-\rho) + \frac{\Lambda}{8\pi}
\mbox{ , }
\label{jan16.L.total}
\end{equation}

For some cosmological models, it is possible to represent $K$ as a time-integral of some quantity.  For example, for the exact cosmological models considered by \cite{EllisMacCallum69}, the effect of $K$ on the action can be represented as an integral over a 4-volume, allowing us to combine the two terms in (\ref{jan16.6.1}).  In that case, the surface term adds the following term to an effective Lagrangian:
\begin{equation}
\frac{3}{2} (p-\rho) - \frac{3\Lambda}{8\pi} - \frac{3 a_0^2 + q_0^2}{4\pi X^2}
\mbox{ , }
\label{jan16.L.surface.eff}
\end{equation}
where $a_0$, $q_0$, and $X$ are parameters of their model, and $a_0$ in (\ref{jan16.L.surface.eff}) has nothing to do with the cosmological scale factor $a_0$ used here.  
If so, then  we would have
\begin{equation}
I = \int (-g^{(4)})^{1/2} 
\tilde{L}
 \mathrm{d}^4x 
\mbox{ , }
\label{jan16.6.1a}
\end{equation}
where $\tilde{L}$ can be considered to be an effective Lagrangian, which in the case of (\ref{jan16.L.total}) and (\ref{jan16.L.surface.eff}) gives
\begin{equation}
\tilde{L} = \alpha(p-\rho) -  \frac{\Lambda}{4\pi}
\mbox{ . }
\label{jan16.L.tilde.special}
\end{equation}
More generally, we can assume the effective Lagrangian to take the form
\begin{equation}
\tilde{L} = \alpha_1 p + \alpha_2 \rho + \alpha_3 \Lambda
\mbox{ , }
\label{jan16.L.tilde}
\end{equation}
where $\alpha_1$, $\alpha_2$, and $\alpha_3$ are dimensionless constants of order unity.  

Table \ref{T3} shows that with vorticity, the density $\rho$ depends on the scale factor along a flow line rather than on the cosmological scale factor $a$.  To take this into account, we can write
\begin{equation}
\tilde{L}(\ell) 
\approx \tilde{L}_0 \left[1+ f_L(a) \delta + f_{LL}(a) \delta^2 \right]
= H_3^2 F (a) \left[1+ f_L(a) \delta + f_{LL}(a) \delta^2 \right]
\mbox{ , }
\label{jan16.L.sec.order}
\end{equation}
where $\delta$ (defined in (\ref{jan16.delta})) is proportional to the square of the present local vorticity,
\begin{eqnarray}
 F(a) =& \frac{3}{8\pi} (\alpha_1 w + \alpha_2) \frac{a_1}{a_2} \left(\frac{a_2}{a}\right)^4  &
\mbox{ for $a_0 \le a \le a_1$, }
\nonumber \\
 F(a) =& \frac{3}{8\pi}  \alpha_2  \left(\frac{a_2}{a}\right)^3 &
\mbox{ for $a_1 \le a \le a_2$, }
\nonumber \\
 F(a) =& 3\alpha_3 &
\mbox{ for $a_2 \le a \le a_3$, }
\label{jan16.F.L}
\end{eqnarray}
\begin{eqnarray}
 f_{L}(a) =& f_\ell(a_1) + 3 f_\ell(a_2) - 4  f_\ell(a) &
\mbox{ for $a_0 \le a \le a_1$, }
\nonumber \\
 f_L(a) =& 3 f_\ell(a_2)  - 3  f_\ell(a) &
\mbox{ for $a_1 \le a \le a_2$, }
\nonumber \\
 f_L(a) =& 0 &
\mbox{ for $a_2 \le a \le a_3$, }
\label{jan16.f.L}
\end{eqnarray}
$f_\ell(a)$ is defined in (\ref{jan16.f.ell.def}),
\begin{eqnarray}
 f_{LL}(a) =& f_{\ell\ell}(a_1) + 3 f_\ell(a_2)^2+ 3 f_{\ell\ell}(a_2) + 10  f_\ell(a)^2- 4  f_{\ell\ell}(a) &
\nonumber \\
&+3 f_\ell(a_1)f_\ell(a_2) -4 f_\ell(a_1)f_\ell(a) -12 f_\ell(a_2)f_\ell(a) &
\mbox{ for $a_0 \le a \le a_1$, }
\nonumber \\
 f_{LL}(a) =& -9 f_\ell(a_2)    f_\ell(a) + 3 f_\ell(a_2)^2+ 3 f_{\ell\ell}(a_2) + 6  f_\ell(a)^2- 3  f_{\ell\ell}(a) &
\mbox{ for $a_1 \le a \le a_2$, }
\nonumber \\
 f_{LL}(a) =& 0 &
\mbox{ for $a_2 \le a \le a_3$, }
\label{jan16.f.LL}
\end{eqnarray}
and $f_{\ell\ell}(a)$ is defined in (\ref{jan16.f.ell.ell}).
Using  (\ref{jan16.f.ell.05}) in (\ref{jan16.f.L}) gives
\begin{eqnarray}
 f_{L}(a) =&& 1 + 3 f_\ell(a_2) - 4  \left(\frac{a}{a_1}\right)^{2}  
\mbox{ for $a_0 \le a \le a_1$, }
\nonumber \\
 f_L(a) =& &
 -216\sqrt{\frac{a_1}{a_2}} +138\frac{a_1}{a_2} +36\frac{a_1}{a_2} \ln{\frac{a_1}{a_2}} +216\sqrt{\frac{a_1}{a}} -138\frac{a_1}{a} -36 \frac{a_1}{a} \ln{\frac{a_1}{a}}
\mbox{ for $a_1 \le a \le a_2$, }
\nonumber \\
&& f_L(a) = 0 
\mbox{ for $a_2 \le a \le a_3$, }
\label{jan16.f.L.2}
\end{eqnarray}

\section{\label{jan16.eras}Radiation, matter, and dark-energy eras}

There are three cosmological eras to consider.  In the early universe, radiation dominates over matter to determine the density $\rho$  in the radical in (\ref{jan16.Hubble-parameter}).  When the cosmological scale factor $a$ reaches a certain size (which we define as $a_1$), matter begins to dominate over radiation to determine the density $\rho$.  When the cosmological scale factor $a$ gets even larger (to a size we define as $a_2$), the density of matter has fallen low enough that the cosmological constant $\Lambda$ begins to dominate over the density term in (\ref{jan16.Hubble-parameter}).  

For an equation of state, we take 
\begin{equation}
p=w\rho
\mbox{ , }
\label{jan16.p.w.rho}
\end{equation}
where $w=1/3$ in the radiation-dominated era, and $w=0$ in the matter-dominated era. The variation of density $\rho$ with cosmological scale factor $a$ is given by \citep[Table 6.1]{Ellis-Maartens-MacCallum:2012} 
\begin{equation}
\rho = \rho_{1} (a/a_{1})^{-3(1+w)} 
\mbox{ , }
\label{jan16.rho.of.r}
\end{equation}
where $\rho_{1}$ is the value of $\rho$ at the boundary between the radiation era and the matter era where $a=a_{1}$.  
Table \ref{T3} summarizes some of this.

We can take for the present fraction of radiation density $\Omega_{\mbox{rad}}=5.4\times 10^{-5}$ \citep[p. 78]{Liddle:2015}.  Otherwise, we take \citep{Planck:Collaboration:2015}
\begin{equation}
a_1 =
 \frac{\Omega_{\mbox{rad}}}{\Omega_{\mbox{mat}}} \approx  \frac{5.4\times 10^{-5}}{.3089} \approx  1.7 \times 10^{-4} 
\mbox{ , }
\label{jan16.r1.ratio}
\end{equation}
and
\begin{equation}
a_2 = 
\left(\frac{\Omega_{\mbox{mat}}}{\Omega_\Lambda}\right)^{1/3} \approx \left(\frac{.3089}{.6911}\right)^{1/3} \approx 0.76 
\mbox{ , }
\label{jan16.r2.ratio}
\end{equation}
where  $\Omega_{\mbox{mat}}$ is the present fraction of matter density (including dark matter), and $\Omega_\Lambda$ is the present fraction of dark energy.   
From the present value of the Hubble parameter \citep{Planck:Collaboration:2015}
\begin{equation}
H_3=67.74 \mbox{ km s$^{-1}$ Mpc$^{-1}$}  =2.195 \times 10^{-18} \mbox{sec$^{-1}$}  =7.323 \times 10^{-29} \mbox{cm$^{-1}$} =6.928 \times 10^{-11} \mbox{yr$^{-1}$}
\mbox{ , }
\label{jan16.H3.numbers}
\end{equation}
we can calculate the critical density, and combining with $\Omega_\Lambda$, we get the value of the density of matter and dark energy when they were equal, which is 
\begin{equation}
\rho_2 = \frac{\Lambda}{8\pi} \approx 4.4\times 10^{-58} \mbox{ cm}^{-2} 
\mbox{ , }
\label{jan16.rho2}
\end{equation}
which gives
\begin{equation}
\Lambda  \approx 1.1 \times 10^{-56} \mbox{ cm}^{-2} 
\mbox{ . }
\label{jan16.Lambda}
\end{equation}
We also have
\begin{equation}
\hbar \to \frac{\hbar G}{c^2} = L^{*2} \approx 2.616  \times 10^{-66} \mbox{ cm}^{2} 
\mbox{ . }
\label{jan16.hbar}
\end{equation}

\section{\label{jan16.relative.rotation}Relative rotation}

It is necessary to estimate the effect of the vorticity $\omega$ on the total rotation angle $\theta$.  Because we expect only very small values of vorticity to contribute significantly to a path integral calculation of vorticity, it is useful to choose a non-dimensional parameter that is small whenever the vorticity is small.  In addition, because the action is an even function of the vorticity, and therefore is a function of the square of vorticity, it is useful for that parameter to be proportional to the square of the vorticity.  It turns out to be useful to take that parameter as

\begin{equation}
\delta \equiv
\frac{1}{6} 
 \left(\frac{\omega_3}{H_3}\right)^2 \left(\frac{a_3}{a_1}\right)
 \left(\frac{a_3}{a_2}\right)^{3} 
\mbox{ ,  }
\label{jan16.delta}
\end{equation}
where $\omega_3$ is the present value of the local vorticity, $H_3$ is the present value of the Hubble parameter, $a_1$ is the value of the cosmological scale factor at the time when the matter density became equal to the radiation density, $a_2$ is the value of cosmological scale factor at the time when the cosmological constant surpassed the matter density in the Friedmann equation, and $a_3=1$ is the present value of the cosmological scale factor.

We start with
\begin{equation}
\theta = \int_0^{t} \omega \, \mathrm{d} t 
= \int_{a_0}^{a} \frac{\omega}{\dot{a}} \, \mathrm{d} a
= \int_{a_0}^{a} \frac{\omega}{aH_0} \left[1+f_H(a) \delta +f_{HH}(a) \delta^2\right] \, \mathrm{d} a
\mbox{ , }
\label{jan16.rotation.a}
\end{equation}
where we have used (\ref{jan16.Friedmann.1d.inv.2}) for $1/\dot{a}$.
Using table \ref{T3} for $\omega$ gives
\begin{eqnarray}
&&
\omega= H_3\sqrt{6\delta} a_1^{-1/2} a_2^{3/2}  (\ell_1/a_1)^{-1}(\ell_3/a_3)^{2}a^{-1}(\ell/a)^{-1}
 \mbox{ for $ a \le a_1$, and}
 \nonumber \\ &&
\omega= H_3\sqrt{6\delta} a_1^{1/2} a_2^{3/2}  
(\ell_3/a_3)^{2}a^{-2}(\ell/a)^{-2}
 \mbox{ for $ a \ge a_1$, }
\label{jan16.omega.1}
\end{eqnarray}
where $\delta$ is defined in (\ref{jan16.delta}).  Using (\ref{jan16.ell.6.second.order}) to expand $\ell_1/a_1$, $\ell_3/a_3$, and $\ell/a$ to second order in $\delta$  allows us to write (\ref{jan16.omega.1}) as
\begin{eqnarray}
&&
\omega= \omega_3 \frac{a_3}{a_1}\frac{a_3}{a}[1+g_\omega \delta + g_{\omega\omega} \delta^2]
 \mbox{ for $ a \le a_1$, and}
 \nonumber \\ &&
\omega= \omega_3 \left(\frac{a_3}{a}\right)^2 [1+g_\omega \delta + g_{\omega\omega} \delta^2]
 \mbox{ for $ a \ge a_1$, }
\label{jan16.omega.2}
\end{eqnarray}
where
\begin{eqnarray}
 g_{\omega}(a) = 
2f_\ell(a_3)-f_\ell(a_1)-f_\ell(a)
  \, \mbox{for } a \le a_1 
 \nonumber \\
 g_{\omega}(a) = 
 2f_\ell(a_3)-2f_\ell(a)
  \, \mbox{for } a\ge a_1
\mbox{ , }
\label{jan16.g.omega}
\end{eqnarray}
$f_\ell(a)$ is defined in (\ref{jan16.f.ell.def}),
\begin{eqnarray}
 g_{\omega\omega}(a) = 
 \left[2 f_{\ell\ell}(a_3) - f_{\ell\ell}(a_1) - f_{\ell\ell}(a) + f_\ell(a_3)^2+ f_\ell(a_1)^2+ f_\ell(a)^2 
 \right.
  \nonumber \\
  \left.
 + f_\ell(a_1)f_\ell(a)-2f_\ell(a_1)f_\ell(a_3)-2f_\ell(a)f_\ell(a_3) \right]
  \, \mbox{for } a \le a_1 
 \nonumber \\
 g_{\omega\omega}(a) = 
 \left[2 f_{\ell\ell}(a_3) -2 f_{\ell\ell}(a) + f_\ell(a_3)^2+ 3f_\ell(a)^2  -4f_\ell(a)f_\ell(a_3) \right]
  \, \mbox{for } a\ge a_1
\mbox{ , }
\label{jan16.g.omega.sec,order}
\end{eqnarray}
and $f_{\ell\ell}(a)$ is defined in (\ref{jan16.f.ell.ell}).

Using (\ref{jan16.omega.1}) for $\omega$, (\ref{jan16.Friedmann.f.H}) for $f_H(a)$, and (\ref{jan16.ell.6.second.order}) to expand $\ell_1/a_1$, $\ell_3/a_3$, and $\ell/a$ to second order in $\delta$  allows us to write (\ref{jan16.rotation.a}) as
\begin{equation}
\theta \approx \sqrt{2 \delta} \left[\sqrt{f_\theta(a)} - \frac{1}{2} f_{\theta \theta} (a)\delta \right]
\mbox{ , }
\label{jan16.rotation.gen.1}
\end{equation}
where
\begin{eqnarray}
&&
\sqrt{f_\theta(a)}= H_3\sqrt{3} a_1^{-1/2} a_2^{3/2}   \int_{a_0}^{a} \frac{a^{-2}}{H_0}  \, \mathrm{d} a
 \mbox{ for $ a \le a_1$, and }
 \nonumber \\ &&
\sqrt{f_\theta(a)}=\sqrt{f_\theta(a_1)}+ H_3\sqrt{3} a_1^{1/2} a_2^{3/2}   \int_{a_1}^{a} \frac{a^{-3}}{H_0}  \, \mathrm{d} a
 \mbox{ for $ a \ge a_1$, }
\label{jan16.f.theta.1}
\end{eqnarray}
and
\begin{eqnarray}
&&
f_{\theta\theta}(a)= -2\left(2f_\ell(a_3) - f_\ell(a_1)\right)\sqrt{f_\theta(a)}
+2 H_3\sqrt{3} a_1^{-1/2} a_2^{3/2}   \int_{a_0}^{a} \frac{a^{-2}}{H_0} (f_\ell(a) - f_H(a)) \, \mathrm{d} a
 \mbox{ for $ a \le a_1$, and }
 \nonumber \\ &&
f_{\theta\theta}(a)=f_{\theta\theta}(a_1) -4f_\ell(a_3)\left(\sqrt{f_\theta(a)}-\sqrt{f_\theta(a_1)}\right)
 \nonumber \\ &&
+2 H_3\sqrt{3} a_1^{1/2} a_2^{3/2}   \int_{a_1}^{a} \frac{a^{-3}}{H_0} (2f_\ell(a) - f_H(a)) \, \mathrm{d} a
 \mbox{ for $ a \ge a_1$. }
\label{jan16.f.theta.theta.new.1}
\end{eqnarray}

Evaluating (\ref{jan16.f.theta.1}) gives
\begin{eqnarray}
&&
f_\theta(a)
= 3
 \left(1 - \frac{a_0}{a} \right)^2
 \left(\frac{a}{a_1}\right)^{2} 
\mbox{ for $a_0 \le a \le a_1$, }
\nonumber \\ &&
f_\theta(a) 
= 3 
 \left[3 -2 \left(\frac{a_1}{a}\right)^{1/2} - \frac{a_0}{a_1} \right]^2
\mbox{ for $a_1 \le a \le a_2$, and }
\nonumber \\ &&
f_\theta(a)
= 3
 \left(\frac{a_1}{a_2}\right) 
 \left[\left(3  - \frac{a_0}{a_1} \right) \left(\frac{a_2}{a_1}\right)^{1/2}
 -\frac{3}{2} -\frac{1}{2} \left(\frac{a_2}{a}\right)^{2} 
 \right]^2
\mbox{ for $a_2 \le a \le a_3$. }
\label{jan16.rotation.8}
\end{eqnarray}

We can neglect $a_0$ in (\ref{jan16.rotation.8}) to give
\begin{eqnarray}
&&
f_\theta(a)
= 3  \left(\frac{a}{a_1}\right)^{2} 
 \mbox{ for $a_0 \le a \le a_1$, }
\nonumber \\ &&
f_\theta(a) 
= 3 
 \left[3 -2 \left(\frac{a_1}{a}\right)^{1/2}  \right]^2
\mbox{ for $a_1 \le a \le a_2$, and }
\nonumber \\ &&
f_\theta(a)
= 3
 \left(\frac{a_1}{a_2}\right) 
  \left[3 \left(\frac{a_2}{a_1}\right)^{1/2}
 -\frac{3}{2} -\frac{1}{2} \left(\frac{a_2}{a}\right)^{2} 
 \right]^2
\mbox{ for $a_2 \le a \le a_3$. }
\label{jan16.rotation.11}
\end{eqnarray}

We also have
\begin{equation}
\theta^2 \approx 2 \delta \left[f_\theta(a) - \sqrt{f_\theta(a)} f_{\theta \theta} (a)\delta \right]
\mbox{ , }
\label{jan16.rotation.gen.2}
\end{equation}
and
\begin{equation}
\cos{\theta} \approx  (1- \frac{1}{2} \theta^2 + \frac{1}{24} \theta^4) 
\approx \left[1-f_\theta(a) \delta +\left(\frac{1}{6}f_\theta^2(a)+\sqrt{f_\theta(a)}f_{\theta\theta}
(a)\right)\delta^2\right]
\mbox{ , }
\label{jan16.cos.theta}
\end{equation}
and
\begin{equation}
\frac{1}{ \cos{\theta} }
\approx  (1+ \frac{1}{2} \theta^2+ \frac{5}{24} \theta^4)
\approx \left[1+f_\theta(a) \delta +\left(\frac{5}{6}f_\theta^2(a)-\sqrt{f_\theta(a)}f_{\theta\theta}(a)\right)\delta^2\right]
\mbox{ , }
\label{jan16.cos.theta.inv}
\end{equation}

Taking the approximate mean of (\ref{jan16.rotation.gen.2}),  using (\ref{jan16.rotation.11}), putting in the appropriate values, including $a_3=1$, gives for the total rms rotation
\begin{equation} \langle\theta\rangle \approx 347 \, \langle\omega_3\rangle /H_3 \approx 347 \, \langle\omega_3\rangle \sqrt{{3}/{\Lambda}} \mbox{ . } \label{jan16.total.rotation.3} \end{equation}
Finally, using the value for $\Lambda$ from (\ref{jan16.Lambda}) and using $\omega_m$ from (\ref{jan16.omegaSub.m}) for $\langle\omega_3\rangle$ gives
\begin{equation} \theta \approx 6 \times 10^{-59} \mbox{ radians }\approx  \times 10^{-58} \mbox{ radians } \label{jan16.total.rotation.4} \end{equation}
for the total rotation of the universe from the initial singularity to the present, showing that it should be valid to assume flow lines are normal to surfaces of constant time for the calculations.  

To find the limits for the validity of the calculation of $f_I(\langle\omega_3\rangle)$, we use (\ref{jan16.total.rotation.3}) to find the value of $\langle\omega_3\rangle$ for which $\theta=1$ to give
 \begin{equation} 
\langle\omega_3\rangle\approx H_3/347 \approx 0.003 H_3
\mbox{ . }
 \label{jan16.total.rotation.5} \end{equation} 
Thus, the approximate calculation of $f_I(\langle\omega_3\rangle)$ is not valid all the way to $\langle\omega_3\rangle=H_3$, but it is valid for $\langle\omega_3\rangle \gg \omega_m$, which is good enough.
Of course this is only the local total rotation.  If we allow for an additional few orders of magnitude increase because of spatial variation of the vorticity, we still have a small total rotation.  

The point is, that it was a good approximation to consider that surfaces of constant global time are normal to flow lines. In addition, there are a few more orders of magnitude to use to allow the approximation to be valid for $\langle\omega_3\rangle$ to be much larger than $\omega_m$ and still be a good approximation.  

It also means that the saddlepoint at $\langle\omega_3\rangle=0$ is isolated from any other saddlepoints or other non-analytic points.

\section{\label{jan16.ell}Approximate global cosmological scale factor}

The flow lines are not normal to surfaces of constant global time, so we have to change variables from $\ell$ to $a$, where $a$ is the cosmological scale factor.  We can use the angle calculations in appendix \ref{jan16.relative.rotation} to make that conversion.  Because the angles are small, we can make approximations.  The formulas below apply to each location in the flow because the vorticity can be a function of location.

We use (\ref{jan16.cos.theta.inv}) 
 to approximate $1/\cos{\theta}$ to give 
\begin{equation}
\ell 
\approx \int \frac{ \, \mathrm{d} a}{\cos{\theta}} 
\approx \int (1+\frac{1}{2} \theta^2+\frac{5}{24} \theta^4) { \, \mathrm{d} a}
\approx \int \left[1+f_\theta(a) \delta +\left(\frac{5}{6}f_\theta^2(a)-\sqrt{f_\theta(a)}f_{\theta\theta}(a)\right)\delta^2\right] { \, \mathrm{d} a}
\mbox{ , }
\label{jan16.ell.0}
\end{equation}
where $a$ is the cosmological scale factor, which is normal to surfaces of constant global time, and $\delta$ is given by (\ref{jan16.delta}).

We can write
\begin{equation}
\ell \approx a\left[1 + f_\ell (a) \delta+ f_{\ell\ell} (a) \delta^2
 \right]
\mbox{ , }
\label{jan16.ell.6.second.order}
\end{equation}

where
\begin{equation}
 f_{\ell} (a) = \frac{1}{a} \int_{a_0}^a 
f_\theta(a)
 \, \mathrm{d} a
\mbox{  }
\label{jan16.f.ell.def}
\end{equation}
and
\begin{equation}
 f_{\ell\ell} (a) = \frac{1}{a} \int_{a_0}^a 
 \left[\frac{5}{6} f_\theta^2(a) -  \sqrt{f_\theta(a)}f_{\theta\theta}(a) \right] 
  \, \mathrm{d} a
\mbox{ . }
\label{jan16.f.ell.ell}
\end{equation}
We also have
\begin{equation}
\frac{1}{\ell} \approx \frac{1}{a}\left[1 - f_\ell (a) \delta+ \left( f_\ell^2 (a)-f_{\ell\ell} (a)\right) \delta^2
 \right]
\mbox{ . }
\label{jan16.ell.inv.6.second.order}
\end{equation}

We can use (\ref{jan16.rotation.11}) in (\ref{jan16.f.ell.def}) and neglect $a_0$ to give
\begin{eqnarray}
&&
f_\ell (a) = 
  \left(\frac{a}{a_1}\right)^{2} 
\mbox{ for $a_0 \le a \le a_1$, }
 \nonumber \\ &&
f_\ell (a) = \left(  27   -72  \sqrt{\frac{a_1}{a}} +46 \frac{a_1}{a} -12  \frac{a_1}{a} \ln{\frac{a_1}{a}}  \right)  
\mbox{ for $a_1 \le a \le a_2$, and }
 \nonumber \\ &&
f_\ell (a) =  
f_\ell (a_2)
\frac{a_2}{a}  + 3 \frac{a_1}{a_2} \left[ 
 \left( 3 \sqrt{\frac{a_2}{a_1}} -\frac{3}{2}\right)^2  \left(1 - \frac{a_2}{a} \right) 
  \right.    \nonumber \\ && \left.  
+\left( 3 \sqrt{\frac{a_2}{a_1}} -\frac{3}{2}\right)  \left(\frac{a_2^2}{a^2} - \frac{a_2}{a} \right)
-\frac{1}{12} \left(\frac{a_2^4}{a^4} - \frac{a_2}{a} \right)   \right] 
\mbox{ for $a_2 \le a \le a_3$. }
\label{jan16.f.ell.05}
\end{eqnarray}

From (\ref{jan16.f.ell.05}), we have
\begin{equation}
f_\ell (a_1) = 1
 \mbox{ . }
\label{jan16.f.ell.a1}
\end{equation}
Using that from (\ref{jan16.r1.ratio}) $a_1$ is so small, we can write
\begin{equation}
f_\ell (a_2) \approx 27
 \mbox{ . }
\label{jan16.f.ell.a2}
\end{equation}

We can write (\ref{jan16.f.ell.ell}) as
\begin{eqnarray}
 f_{\ell\ell} (a) = \frac{1}{a} \int_{a_0}^a 
 \left[\frac{5}{6} f_\theta^2(a) -  \sqrt{f_\theta(a)}f_{\theta\theta}(a) \right] 
 \, \mathrm{d} a
\mbox{ for $a_0 \le a \le a_1$,  }
  \nonumber \\  
 f_{\ell\ell} (a) = 
 f_{\ell\ell} (a_1) \frac{a_1}{a}
 +
  \frac{1}{a} \int_{a_1}^a
 \left[\frac{5}{6} f_\theta^2(a) -  \sqrt{f_\theta(a)}f_{\theta\theta}(a) \right] 
    \, \mathrm{d} a
\mbox{ for $a_1 \le a \le a_2$,  }
  \nonumber \\  
 f_{\ell\ell} (a) = 
 f_{\ell\ell} (a_2) \frac{a_2}{a}
 +
  \frac{1}{a} \int_{a_2}^a 
 \left[\frac{5}{6} f_\theta^2(a) -  \sqrt{f_\theta(a)}f_{\theta\theta}(a) \right] 
   \, \mathrm{d} a
\mbox{ for $a_2 \le a \le a_3$,  }
\label{jan16.f.ell.ell.2}
\end{eqnarray}
Or,
\begin{eqnarray}
 f_{\ell\ell} (a) =  
 \frac{5}{6} \frac{1}{a} \int_{a_0}^a \left(f_\theta(a)\right)^2  \, \mathrm{d} a - \frac{1}{a} \int_{a_0}^a f_{\theta\theta}(a) \sqrt{f_\theta(a)}  \, \mathrm{d} a
\mbox{ for $a_0 \le a \le a_1$,  }
  \nonumber \\  
 f_{\ell\ell} (a) = 
 f_{\ell\ell} (a_1) \frac{a_1}{a}
  +
  \frac{5}{6}\frac{1}{a} \int_{a_1}^a  \left(f_\theta(a)\right)^2  \, \mathrm{d} a - \frac{1}{a} \int_{a_1}^a f_{\theta\theta}(a) \sqrt{f_\theta(a)}   \, \mathrm{d} a
\mbox{ for $a_1 \le a \le a_2$,  }
  \nonumber \\  
 f_{\ell\ell} (a) = 
  f_{\ell\ell} (a_2) \frac{a_2}{a}
 +
  \frac{5}{6}  \frac{1}{a} \int_{a_2}^a \left(f_\theta(a)\right)^2 \, \mathrm{d} a - \frac{1}{a} \int_{a_2}^a  f_{\theta\theta}(a) \sqrt{f_\theta(a)} \, \mathrm{d} a 
\mbox{ for $a_2 \le a \le a_3$,  }
\label{jan16.f.ell.ell.3}
\end{eqnarray}

\input{jan16.AppF-Vorticity}

\section{\label{jan16.main}Appproximate action for small vorticity}

The main effect of vorticity and shear on the action is through the generalized Friedmann equation (\ref{jan16.Friedmann.1.ell}).  Starting with (\ref{jan16.6.1a}), change the $t$ integration to an integration over $a$ using (\ref{jan16.Friedmann.1d.inv.2}) and use (\ref{jan16.L.sec.order}) for $\tilde{L}$ to give
\begin{equation}
I 
= \int
 \frac{  V \tilde{L}  \, \mathrm{d} \, a }{\dot{a}}
\approx \int_{a_0}^{a_3}
 \frac{  V (a)  H_3^2 F(a) }{ a H_0} \left[1+ f_L(a) \overline{\delta} + f_{LL}(a) \overline{\delta^2} \right] \left[1+f_H(a) \overline{\delta} +f_{HH}(a) \overline{\delta^2}\right]\, \mathrm{d} \, a
\mbox{ , }
\label{jan16.6.1.c.1}
\end{equation}
where 
\begin{equation} V(a) = \frac{4}{3} \pi a^3 r_3^3 \mbox{  } \label{jan16.volume} \end{equation} 
is the approximate spatial volume\footnote{Since our universe is approximately spatially flat.  Although there are arguments that $V$ should be infinite because this is an open cosmology \citep{Hartle2005}, considering causality leads to restricting the spatial part of the action to the past light cone.  Besides, an infinite action would give the result that the only cosmologies contributing significantly to the path integral had a vorticity exactly equal to zero.}, $a$ is the cosmological scale factor, $r_3$ is the present radius of the cosmological horizon,  
\begin{equation}
\overline{\delta} =
\frac{1}{6} 
 \overline{\left(\frac{\omega_3}{H_3}\right)^2} \left(\frac{a_3}{a_1}\right)
 \left(\frac{a_3}{a_2}\right)^{3} 
=
\frac{1}{6} 
 \left(\frac{\langle\omega_3\rangle}{H_3}\right)^2 \left(\frac{a_3}{a_1}\right)
 \left(\frac{a_3}{a_2}\right)^{3} 
 \mbox{   }
\label{jan16.delta.average}
\end{equation}
is the spatial average of $\delta$ given by (\ref{jan16.delta}), 
\begin{equation}
\overline{\delta^2} =
\overline{\delta}^2 +
\left[\frac{1}{6} 
 \left(\frac{1}{H_3}\right)^2 \left(\frac{a_3}{a_1}\right)
 \left(\frac{a_3}{a_2}\right)^{3} \right]^2 \sigma_\omega^2
 \mbox{   }
\label{jan16.delta.squared.average}
\end{equation}
is the spatial average of the square of $\delta$,  $\sigma_\omega^2$ is the variance of $\omega_3^2$, and we have neglected shear and acceleration, keeping only vorticity.  Multiplying out, and keeping terms only up to second order in $\overline{\delta}$ gives
\begin{equation}
I \approx
 \int_{a_0}^{a_3}
 \frac{  V (a)  H_3^2 F(a) }{ a H_0} \left\{1+ \left[f_L(a)+f_H(a) \right] \overline{\delta} + 
 \left[f_L(a)f_H(a) + f_{LL}(a) +f_{HH}(a)\right]\overline{\delta^2} \right\} 
 \, \mathrm{d} \, a
\mbox{ . }
\label{jan16.6.1.c}
\end{equation}

We can write (\ref{jan16.6.1.c}) as
\begin{equation}
I \approx I_0 + I_1 \overline{\delta} + I_2 \overline{\delta^2} 
\mbox{ , }
\label{jan16.6.1.c.sec.order}
\end{equation}
where $I_0$ is the action for zero vorticity and zero shear, 
\begin{equation}
I_1 
= \int_{a_0}^{a_3}  \frac{  V (a)  H_3^2 F(a) }{ a H_0}
\left[f_L(a)+f_H(a) \right]
\, \mathrm{d} \, a
\mbox{ . }
\label{jan16.6.I.1}
\end{equation}
and
\begin{equation}
I_2 
= \int_{a_0}^{a_3} \frac{  V (a)  H_3^2 F(a) }{ a H_0}
\left[f_L(a)f_H(a) + f_{LL}(a) +f_{HH}(a)\right]
\, \mathrm{d} \, a
\mbox{ . }
\label{jan16.6.I.2}
\end{equation}

We can use (\ref{jan16.volume}) for $V(a)$,  and (\ref{jan16.delta.average}) for $\overline{\delta}$  and (\ref{jan16.delta.squared.average}) for $\overline{\delta^2}$ to write (\ref{jan16.6.1.c.sec.order}) as
\begin{equation}
I \approx I_0 + \hbar \left(\frac{\langle\omega_3\rangle}{\omega_m}\right)^2 
\left(\frac{a_3}{a_1}\right)
\left[C_I + \frac{\langle\omega_3\rangle^2+\sigma_\omega^2/\langle\omega_3\rangle^2}{H_3^2}
\left(\frac{a_3}{a_1}\right)
C_{II}\right]
\mbox{ , }
\label{jan16.6.1.c.sec.order.2}
\end{equation}
where 
\begin{equation}
 \omega_m
= 
 \left( \frac{ \hbar 
H_3
}{ r_3^3 } \right)^{1/2}
=  \frac{T^*}{r_3}\sqrt{\frac{H_3}{r_3}}
\approx  T^* H_3^2
\approx 
10^{-89} \mbox{cm}^{-1} 
 \approx    10^{-71} \mbox{rad yr$^{-1}$}
\mbox{ , }
\label{jan16.omegaSub.m}
\end{equation}
$H_3$ is the present value of the Hubble parameter, $r_3$ is the present radius of the universe (which we approximate by the inverse of the Hubble parameter), and $T^*$ is the Planck time,
\begin{equation}
C_I = 
\frac{1}{6} 
 \left(\frac{a_3}{a_2}\right)^{3} 
 \frac{4}{3} \pi  
 \int_{a_0}^{a_3} a^2 \frac{   H_3 F (a)}{  H_0 } 
\left[f_L(a)+f_H(a) \right]
\, \mathrm{d} \, a
\mbox{ , }
\label{jan16.6.f.I.0}
\end{equation}
and
\begin{equation}
C_{II} 
= 
\frac{1}{36} 
 \left(\frac{a_3}{a_2}\right)^{6} 
 \frac{4}{3} \pi  
 \int_{a_0}^{a_3} a^2\frac{   H_3 F (a)}{  H_0 } 
 \left[f_L(a)f_H(a) + f_{LL}(a) +f_{HH}(a)\right]
\, \mathrm{d} \, a
\mbox{ . }
\label{jan16.6.f.II.0}
\end{equation}

We can use (\ref{jan16.Friedmann.f.H}) for $f_H(a)$  in (\ref{jan16.6.f.I.0}) to give
\begin{equation}
C_I = 
\frac{1}{6} 
 \left(\frac{a_3}{a_2}\right)^{3} 
 \frac{4}{3} \pi  
 \int_{a_0}^{a_3} a^2 \frac{   H_3 F (a)}{  H_0 } 
\left[f_\theta(a) - f_\ell(a)+\frac{1}{2}f_L(a)  - \frac{1}{2} \frac{H_3^2}{H_0^2}f_\omega(a)\right] 
\, \mathrm{d} \, a
\mbox{ , }
\label{jan16.6.f.I}
\end{equation}

We can use (\ref{jan16.Friedmann.f.H}) for $f_H(a)$ and (\ref{jan16.Friedmann.f.HH}) for $f_{HH}(a)$ in (\ref{jan16.6.f.II.0}) to give
\begin{eqnarray}
C_{II} 
= 
\frac{1}{36} 
 \left(\frac{a_3}{a_2}\right)^{6} 
 \frac{4}{3} \pi  
 \int_{a_0}^{a_3} a^2\frac{   H_3 F (a)}{  H_0 } 
\left\{ f_\ell(a)^2- f_{\ell\ell} (a) -f_\theta(a) f_\ell(a) 
 \right.
  \nonumber \\
+\frac{5}{6} f_\theta(a)^2 -  \sqrt{f_\theta(a)} f_{\theta\theta}(a)  
  + \frac{1}{2}f_{LL}(a) + 
  \frac{1}{2}f_{L}(a)f_\theta(a) - \frac{1}{2}f_{L}(a)f_\ell(a) -
  \frac{1}{8}f_{L}^2(a)
  \nonumber \\
  \left.
 - \frac{1}{2} \frac{H_3^2}{H_0^2} \left[ 
 f_\omega(a)f_\theta(a) - f_\omega(a)f_\ell(a)
  -\frac{1}{2}f_L(a) f_\omega(a)
 + f_{\omega\omega}(a) - \frac{3}{4} f_\omega(a)^2 \frac{H_3^2}{H_0^2} \right]\right\} 
\, \mathrm{d} \, a
\mbox{ . }
\label{jan16.6.f.II}
\end{eqnarray}

We can write (\ref{jan16.6.f.I}) as 
\begin{eqnarray}
C_I 
= 
 \left(\frac{a_3}{a_2}\right)^{3} 
 \frac{2\pi }{9}  
 \left\{
 \int_{a_0}^{a_1} a^2 \frac{  
H_3 
F (a)
}{  H_0 } 
\left[
f_\theta(a) 
- f_\ell(a)
+\frac{1}{2}f_L(a)
 - \frac{1}{2} \frac{H_3^2}{H_0^2}
 f_\omega(a)
\right] 
\, \mathrm{d} \, a
\right.
  \nonumber \\
  +
   \int_{a_1}^{a_2} a^2 \frac{  
H_3 
F (a)
}{  H_0 } 
\left[
f_\theta(a) 
- f_\ell(a)
+\frac{1}{2}f_L(a)
 - \frac{1}{2} \frac{H_3^2}{H_0^2}
 f_\omega(a)
\right] 
\, \mathrm{d} \, a
  \nonumber \\
  \left.
  +
   \int_{a_2}^{a_3} a^2 \frac{  
H_3 
F (a)
}{  H_0 } 
\left[
f_\theta(a) 
- f_\ell(a)
+\frac{1}{2}f_L(a)
 - \frac{1}{2} \frac{H_3^2}{H_0^2}
 f_\omega(a)
\right] 
\, \mathrm{d} \, a
\right\}
\mbox{ . }
\label{jan16.6.f.I.2}
\end{eqnarray}
We can use 
(\ref{jan16.Hubble-parameter.0}) for $H_0$, 
(\ref{jan16.F.L}) for $ F (a)$, 
(\ref{jan16.rotation.11}) for $f_\theta(a)$, 
(\ref{jan16.f.ell.05}) for $f_\ell(a)$, 
(\ref{jan16.f.L.2}) for $f_L(a)$, 
and 
(\ref{jan16.H.f1}) for $f_\omega(a)$, 
in (\ref{jan16.6.f.I.2}).  
Keeping only the dominant term in the Hubble factor in each era, using the appropriate dependence on $a$ of $\rho_0$ in each era, performing the integrations, neglecting $a_0$ when appropriate, using (\ref{jan16.r1.ratio}) and (\ref{jan16.r2.ratio}) to allow us to neglect $a_1$ in some places, gives

\begin{eqnarray}
&
C_I 
= 
  {w}
 \left(\frac{a_1}{a_2}\right)^{3/2}
 {a_3^3}
\left[
\frac{41}{12}
 - \frac{1}{18}  \alpha_4 
 \right] \alpha_1  
  \nonumber \\
  &
  +
  {a_3^3}
\left[  6 \left(\frac{a_1}{a_2}\right)^{1/2}
-
\frac{1}{15} 
 \frac{a_2}{a_1}
 \alpha_4
\right] \alpha_2
  \nonumber \\
  &
  +  {2\pi }
   {a_3^3}
  \left(\frac{a_3}{a_2}- 1\right)
\left[ 
 \frac{9}{2}
  \left(\frac{a_3}{a_2}+1\right)
 - 
 \frac{2 }{3} 
\frac{a_2}{a_1}
\alpha_4
\right] \alpha_3 
\mbox{ . }
\label{jan16.6.f.I.7d}
\end{eqnarray}

Putting in values for $w$, $a_1$, $a_2$, and $a_3=1$ from appendix  \ref{jan16.eras} gives
\begin{eqnarray}
&
C_I 
= 
 \left(
 4\times 10^{-6}
 -6\times 10^{-8}  \alpha_4 
\right) \alpha_1  
  +
\left(  
9\times 10^{-2}
-
298
 \alpha_4
\right) \alpha_2
  \nonumber \\   &
  + 
  2\pi\left( 
3.3
 - 
941
\alpha_4
\right) \alpha_3 
\mbox{ . }
\label{jan16.6.f.I.7e}
\end{eqnarray}

We can write (\ref{jan16.6.f.II}) as
\begin{eqnarray}
C_{II} 
= 
\frac{1}{36} 
 \left(\frac{a_3}{a_2}\right)^{6} 
 \frac{4}{3} \pi  
 \int_{a_0}^{a_1} a^2\frac{   H_3 F (a)}{  H_0 } 
\left\{ f_\ell(a)^2- f_{\ell\ell} (a) -f_\theta(a) f_\ell(a) 
 \right.
  \nonumber \\
+\frac{5}{6} f_\theta(a)^2 -  \sqrt{f_\theta(a)} f_{\theta\theta}(a)  
  + \frac{1}{2}f_{LL}(a) + 
  \frac{1}{2}f_{L}(a)f_\theta(a) - \frac{1}{2}f_{L}(a)f_\ell(a) -
  \frac{1}{8}f_{L}^2(a)
  \nonumber \\
  \left.
 - \frac{1}{2} \frac{H_3^2}{H_0^2} \left[ 
 f_\omega(a)f_\theta(a) - f_\omega(a)f_\ell(a)
  -\frac{1}{2}f_L(a) f_\omega(a)
 + f_{\omega\omega}(a) - \frac{3}{4} f_\omega(a)^2 \frac{H_3^2}{H_0^2} \right]\right\} 
\, \mathrm{d} \, a
  \nonumber \\ +
\frac{1}{36} 
 \left(\frac{a_3}{a_2}\right)^{6} 
 \frac{4}{3} \pi  
 \int_{a_1}^{a_2} a^2\frac{   H_3 F (a)}{  H_0 } 
\left\{ f_\ell(a)^2- f_{\ell\ell} (a) -f_\theta(a) f_\ell(a) 
 \right.
  \nonumber \\
+\frac{5}{6} f_\theta(a)^2 -  \sqrt{f_\theta(a)} f_{\theta\theta}(a)  
  + \frac{1}{2}f_{LL}(a) + 
  \frac{1}{2}f_{L}(a)f_\theta(a) - \frac{1}{2}f_{L}(a)f_\ell(a) -
  \frac{1}{8}f_{L}^2(a)
  \nonumber \\
  \left.
 - \frac{1}{2} \frac{H_3^2}{H_0^2} \left[ 
 f_\omega(a)f_\theta(a) - f_\omega(a)f_\ell(a)
  -\frac{1}{2}f_L(a) f_\omega(a)
 + f_{\omega\omega}(a) - \frac{3}{4} f_\omega(a)^2 \frac{H_3^2}{H_0^2} \right]\right\} 
\, \mathrm{d} \, a
  \nonumber \\
+
\frac{1}{36} 
 \left(\frac{a_3}{a_2}\right)^{6} 
 \frac{4}{3} \pi  
 \int_{a_2}^{a_3} a^2\frac{   H_3 F (a)}{  H_0 } 
\left\{ f_\ell(a)^2- f_{\ell\ell} (a) -f_\theta(a) f_\ell(a) 
 \right.
  \nonumber \\
+\frac{5}{6} f_\theta(a)^2 -  \sqrt{f_\theta(a)} f_{\theta\theta}(a)  
  + \frac{1}{2}f_{LL}(a) + 
  \frac{1}{2}f_{L}(a)f_\theta(a) - \frac{1}{2}f_{L}(a)f_\ell(a) -
  \frac{1}{8}f_{L}^2(a)
  \nonumber \\
  \left.
 - \frac{1}{2} \frac{H_3^2}{H_0^2} \left[ 
 f_\omega(a)f_\theta(a) - f_\omega(a)f_\ell(a)
  -\frac{1}{2}f_L(a) f_\omega(a)
 + f_{\omega\omega}(a) - \frac{3}{4} f_\omega(a)^2 \frac{H_3^2}{H_0^2} \right]\right\} 
\, \mathrm{d} \, a
\mbox{ . }\label{jan16.6.f.II.2}\end{eqnarray}

We can use (\ref{jan16.Hubble-parameter.0}) for $H_0$, keep only the dominant term in the Hubble factor in each era, use (\ref{jan16.H3}) for $H_3$, use the appropriate dependence on $a$ of $\rho_0$ in each era, and use (\ref{jan16.F.L}) for $ F (a)$ to give 
\begin{eqnarray}
C_{II} 
= 
\frac{\alpha_1 w+\alpha_2}{72} a_1^{1/2}a_2^{-9/2} a_3^6
 \int_{a_0}^{a_1} 
  \left\{ f_\ell(a)^2- f_{\ell\ell} (a) -f_\theta(a) f_\ell(a) 
 \right.
  \nonumber \\
+\frac{5}{6} f_\theta(a)^2 -  \sqrt{f_\theta(a)} f_{\theta\theta}(a)  
  + \frac{1}{2}f_{LL}(a) + 
  \frac{1}{2}f_{L}(a)f_\theta(a) - \frac{1}{2}f_{L}(a)f_\ell(a) -
  \frac{1}{8}f_{L}^2(a)
  \nonumber \\
  \left.
 - \frac{1}{2} \frac{H_3^2}{H_0^2} \left[ 
 f_\omega(a)f_\theta(a) - f_\omega(a)f_\ell(a)
  -\frac{1}{2}f_L(a) f_\omega(a)
 + f_{\omega\omega}(a) - \frac{3}{4} f_\omega(a)^2 \frac{H_3^2}{H_0^2} \right]\right\} 
\, \mathrm{d} \, a
  \nonumber \\ +
\frac{\alpha_2}{72} a_2^{-9/2} a_3^6
 \int_{a_1}^{a_2} a^{1/2}
\left\{ f_\ell(a)^2- f_{\ell\ell} (a) -f_\theta(a) f_\ell(a) 
 \right.
  \nonumber \\
+\frac{5}{6} f_\theta(a)^2 -  \sqrt{f_\theta(a)} f_{\theta\theta}(a)  
  + \frac{1}{2}f_{LL}(a) + 
  \frac{1}{2}f_{L}(a)f_\theta(a) - \frac{1}{2}f_{L}(a)f_\ell(a) -
  \frac{1}{8}f_{L}^2(a)
  \nonumber \\
  \left.
 - \frac{1}{2} \frac{H_3^2}{H_0^2} \left[ 
 f_\omega(a)f_\theta(a) - f_\omega(a)f_\ell(a)
  -\frac{1}{2}f_L(a) f_\omega(a)
 + f_{\omega\omega}(a) - \frac{3}{4} f_\omega(a)^2 \frac{H_3^2}{H_0^2} \right]\right\} 
\, \mathrm{d} \, a
  \nonumber \\
+
\frac{\pi\alpha_3}{9} 
 \left(\frac{a_3}{a_2}\right)^{6} 
 \int_{a_2}^{a_3} a^2
\left\{ f_\ell(a)^2- f_{\ell\ell} (a) -f_\theta(a) f_\ell(a) 
 \right.
  \nonumber \\
+\frac{5}{6} f_\theta(a)^2 -  \sqrt{f_\theta(a)} f_{\theta\theta}(a)  
  + \frac{1}{2}f_{LL}(a) + 
  \frac{1}{2}f_{L}(a)f_\theta(a) - \frac{1}{2}f_{L}(a)f_\ell(a) -
  \frac{1}{8}f_{L}^2(a)
  \nonumber \\
  \left.
 - \frac{1}{2} \frac{H_3^2}{H_0^2} \left[ 
 f_\omega(a)f_\theta(a) - f_\omega(a)f_\ell(a)
  -\frac{1}{2}f_L(a) f_\omega(a)
 + f_{\omega\omega}(a) - \frac{3}{4} f_\omega(a)^2 \frac{H_3^2}{H_0^2} \right]\right\} 
\, \mathrm{d} \, a
\mbox{ . }
\label{jan16.6.f.II.3}
\end{eqnarray}

We can use  (\ref{jan16.rotation.11}) for $f_\theta(a)$,  (\ref{jan16.f.ell.05}) for $f_\ell(a)$, 
(\ref{jan16.f.L.2}) for $f_L(a)$, 
(\ref{jan16.H.f1}) for $f_\omega(a)$, 
(\ref{jan16.f.ell.ell.3}) for $f_{\ell\ell} (a)$, 
(\ref{jan16.f.theta.theta.new.1})  for $f_{\theta\theta}(a) $, 
(\ref{jan16.f.LL}) for $f_{LL}(a)$, 
(\ref{jan16.H.f1.sec.order.2}) for $f_{\omega\omega}(a)$, 
neglect $a_0$, 
and use the fact that $a_1/a_2 \approx 10^{-4}$ to approximate (\ref{jan16.6.f.II.3}) as 
\begin{eqnarray}
&&C_{II} 
\approx
 \frac{1}{72} 
\left(\frac{a_1}{a_2}\right)^{3/2}\left(\frac{a_3}{a_2}\right)^{3}a_3^{3}
\left[ 
\frac{52007}{9000} -\frac{5}{12}f_\ell(a_2)+\frac{40}{9}f_\ell(a_3)+\frac{1}{2}f_{\ell\ell}(a_1)+\frac{3}{2}f_{\ell\ell}(a_2)
\right.
  \nonumber \\ &&
  \left.
  +\frac{3}{8}f_\ell^2(a_2)+\frac{3}{2}f_\ell(a_1)f_\ell(a_2)+\left(-\frac{211}{75}+f_\ell(a_2)-\frac{8}{3}f_\ell(a_3)\right)\alpha_4+\frac{6}{5}\alpha_4^2
\right] (\alpha_1 w + \alpha_2)
  \nonumber \\ &&
+ \frac{1}{72} 
\left(\frac{a_2}{a_1}\right)^{1/2}\left(\frac{a_3}{a_2}\right)^{3}a_3^{3}
 \left[ 
-144
 -\frac{208}{5}\sqrt{\frac{a_2}{a_1}}\alpha_4 +\frac{12}{7}  \frac{a_2}{a_1} \alpha_4^2
\right] \alpha_2 
  \nonumber \\ &&
+ 
\frac{ 2\pi }{3} \left(\frac{a_3}{a_2}\right)^{3} a_3^{3} \left(1-\frac{a_2}{a_3}\right)
\left[ 
 \frac{3}{2} \sqrt{\frac{a_2}{a_1}} 
 -\frac{1}{3}  \frac{a_3}{a_1}   \alpha_4 
 +  \left( \frac{a_2}{a_1}\right)^2  \alpha_4^2 
 \right] \alpha_3
\mbox{ . }
\label{jan16.6.f.II.4}
\end{eqnarray}
Or,
\begin{eqnarray}
&&C_{II} 
\approx
 \frac{1}{72} 
\left(\frac{a_1}{a_2}\right)^{3/2}\left(\frac{a_3}{a_2}\right)^{3}a_3^{3}
\left[ 1066
-72 \sqrt{\frac{a_2}{a_1}}
\alpha_4+\frac{6}{5}\alpha_4^2
\right] (\alpha_1 w + \alpha_2)
  \nonumber \\ &&
+ \frac{1}{72} 
\left(\frac{a_2}{a_1}\right)^{1/2}\left(\frac{a_3}{a_2}\right)^{3}a_3^{3}
 \left[ 
-144
 -\frac{208}{5}\sqrt{\frac{a_2}{a_1}}\alpha_4 +\frac{12}{7}  \frac{a_2}{a_1} \alpha_4^2
\right] \alpha_2 
  \nonumber \\ &&
+ 
\frac{ 2\pi }{3} \left(\frac{a_3}{a_2}\right)^{3} a_3^{3} \left(1-\frac{a_2}{a_3}\right)
\left[ 
 \frac{3}{2} \sqrt{\frac{a_2}{a_1}} 
 -\frac{1}{3}  \frac{a_3}{a_1}   \alpha_4 
 +  \left( \frac{a_2}{a_1}\right)^2  \alpha_4^2 
 \right] \alpha_3
\mbox{ . }
\label{jan16.6.f.II.4a}
\end{eqnarray}

Putting in values for $w$, $a_1$, $a_2$, and $a_3=1$ from appendix  \ref{jan16.eras} gives
\begin{eqnarray}
&&C_{II} 
\approx
   \left(4
    \times 
   10^{-5}    -2      \times 10^{-4}  {\alpha_4 } +4      \times 10^{-8}  {\alpha_4 ^2}\right)
\alpha_1 
  \nonumber \\ &&
+
\left( -3\times 10^{2} -6\times10^{3}    \alpha_4 +2     \times 10^{4}  {\alpha_4 ^2} \right) \alpha_2 
 \nonumber \\ &&
+  8\pi\left( 
5 -10^2   \alpha_4 +10^6   \alpha_4^2 
\right) \alpha_3
\mbox{ . }
\label{jan16.6.f.II.5}
\end{eqnarray}

%% file: jan16.AppF-Vorticity.tex
\section{\label{jan16.Friedmann}Approximate Generalized Friedmann equation for small Vorticity}

It is possible to find an approximate solution to the field equations that is a perturbation from the standard cosmological model for small vorticity.  The difficulty with finding an approximate solution that includes vorticity is that the Raychaudhuri equation is based on a coordinate system that follows flow lines, but surfaces of constant global time cannot be found that are normal to the flow lines.  However, in the limit of small vorticity, we can estimate the effect that the scale factor along flow lines $\ell$ is not quite the same as the global cosmological scale factor $a$.

Derivation of a generalization of the Friedmann equation that includes relative rotation of matter and inertial frames (specifically, shear and vorticity), starts with the Raychaudhuri 
equation \citep{Raychaudhuri:1955,Raychaudhuri:1957}, \citep[eq. 1.3.4]{Ehlers:1961}, \citep[eq. (36)]{Ehlers:1993}, \citep[eq. 4.12]{Ellis:1971,Ellis:2009}, \citep{Kubo:1978,Kramer-Stephani-MacCallum-Herlt:1980,Stephani-Kramer-MacCallum-Hoenselaers-Herlt:2003,Ellis-Siklos-Wainwright:1997,Szydlowski-Godlowsk-Flin-Biernacka:2003,Szydlowski-Godlowsk:2003}, or the 
Raychaudhuri-Ehlers equation \citep[eq. 6.4]{Ellis-Maartens-MacCallum:2012}.

We start with the Raychaudhuri-Ehlers equation \citep[eq. 6.5]{Ellis-Maartens-MacCallum:2012} 
\begin{equation}
3 \frac{\ddot{\ell}}{\ell}
 = -2\left(
   \sigma^2
  - \omega^2\right)
    + {\nabla}_a\dot{u}^a + \dot{u}_a\dot{u}^a - 4\pi G (\rho+3p) + \Lambda
 \mbox{ , }
\label{jan16.Raychaudhuri.1}
\end{equation}
where $\ell$ is a scale factor that follows flow lines, and $\cdot$ is a derivative with respect to a variable $\tau$ that increases along the flow line $u$.

Multiplying (\ref{jan16.Raychaudhuri.1}) by $\ell \dot{\ell}$ and integrating gives
\begin{equation}
{\dot{\ell}}/{\ell}
 =  \sqrt{ H^2 
  + H_{\omega}^2 + H_{\sigma}^2 + H_{a}^2 }
\mbox{ , }
\label{jan16.Friedmann.1.ell}
\end{equation}
where
\begin{equation}
H \equiv \sqrt{\frac{\Lambda}{3}  + \frac{8\pi \rho}{3}   
 - \frac{k}{{\ell}^2}   }
\mbox{  }
\label{jan16.Hubble-parameter.ell}
\end{equation}
is the Hubble parameter without vorticity, shear, or acceleration,
\begin{equation} H_{\omega}^2 \equiv \frac{4}{3\ell^2}    \int \ell  \omega^2  \, \mathrm{d} \ell \mbox{  } \label{jan16.H.omega.ell} \end{equation}
is the vorticity term, $\omega$ is vorticity, 
\begin{equation} H_{\sigma}^2 \equiv -\frac{4}{3\ell^2}    \int \ell  \sigma^2  \, \mathrm{d} \ell \mbox{  } \label{jan16.H.sigma.ell} \end{equation}
is the shear term, $\sigma$ is shear, and 
\begin{equation} H_{a}^2 \equiv - \frac{2}{3\ell^2} \int \ell {\dot{u}^a}_{\hspace{4pt};a}  \, \mathrm{d} \ell  \mbox{  } \label{jan16.H.a.ell} \end{equation}
is the acceleration term. 

We can take vorticity to depend on the distance along flow lines $\ell$ as 
 \begin{equation}
\omega \propto {\ell}^{-m}
\mbox{ , }
\label{jan16.vorticity.1}
\end{equation}
and, for small vorticity, we can take $m=1$ in the radiation era and $m=2$ in the matter era \citep[Table 6.1]{Ellis-Maartens-MacCallum:2012}.\footnote{Instead of this simple assumption, we could instead take into account the coupling between vorticity and shear in a combined way in terms of a vector perturbation \citep[Chapter 10]{Ellis-Maartens-MacCallum:2012}.  The final effect on the calculation of the total action, however, is not significant.}
If we put the $\ell$-variation of vorticity into (\ref{jan16.H.omega.ell}), then we get 
\begin{equation}
H_{\omega}^2 \approx H_3^2 f_{\omega}(a) \delta 
\mbox{ , }
\label{jan16.H.omega.2}
\end{equation}
where $\delta$ is given by (\ref{jan16.delta}), 
\begin{equation}
H_{3} \approx \sqrt{\Lambda/3}
\mbox{  }
\label{jan16.H3}
\end{equation}
is the present value of the Hubble parameter, and 
\begin{eqnarray}
 f_{\omega}(a) = 
 4   \left(\frac{a_2}{a_1}\right)^3 \left(\frac{a_1}{a}\right)^2
 \left[2  \ln{\frac{a}{a_1}}   +  \alpha_4 -1
 \right]
  \, \mbox{for } a \le a_1 
 \nonumber \\
 f_{\omega}(a) = 
 4   \left(\frac{a_2}{a_1}\right)^3 \left(\frac{a_1}{a}\right)^2
\left[\alpha_4 - \left(\frac{a_1}{ a}\right)^2 \right]
  \, \mbox{for } a\ge a_1
\mbox{ , }
\label{jan16.H.f1}
\end{eqnarray}
where $\alpha_4$ is an arbitrary constant of integration in (\ref{jan16.H.omega.ell}).  

We can generalize (\ref{jan16.H.omega.2}) to second order by writing
\begin{equation}
H_{\omega}^2 = H_3^2 \left[f_{\omega}(a)  \delta + f_{\omega\omega}(a)  \delta^2 \right]
\mbox{ , }
\label{jan16.H.omega.2.sec.order}
\end{equation}
where
\begin{eqnarray}
 f_{\omega\omega}(a) = 
 8   \left(\frac{a_2}{a_1}\right)^3 \left(\frac{a_1}{a}\right)^2 
 \left[\left(2 f_\ell(a_3) - f_\ell(a_1) - f_\ell(a)\right)\left(2\ln{\frac{a}{a_1}}+\alpha_4-1\right) + f_\ell(a) - f_\ell(a_1) \right]
  \nonumber \\
  \, \mbox{for } a \le a_1 
 \nonumber \\
 f_{\omega\omega}(a) = 
 8   \left(\frac{a_2}{a_1}\right)^3 \left(\frac{a_1}{a}\right)^2 
\left[\left(2f_\ell(a_3) - f_\ell(a_1)- f_\ell(a)\right)
\alpha_4 
 -2 \left(f_\ell(a_3) - f_\ell(a)\right) \left(\frac{a_1}{ a}\right)^2\right]
  \nonumber \\
   \, \mbox{ for } a\ge a_1 
\mbox{ . } \,\,\,\,\,\,\,\,\,\,
\label{jan16.H.f1.sec.order}
\end{eqnarray}

We can use (\ref{jan16.f.ell.05}) for $f_\ell(a)$ in (\ref{jan16.H.f1.sec.order}) to give 
\begin{eqnarray}
 f_{\omega\omega}(a) = 
 8   \left(\frac{a_2}{a_1}\right)^3 
 \left[2 \left(2 f_\ell(a_3) - 1 
  \right)\left(\frac{a_1}{a}\right)^2 \ln{\frac{a}{a_1}} 
  -2\ln{\frac{a}{a_1}} 
 \right.
  \nonumber \\
  \left.
  + (2 f_\ell(a_3) - 1)  (\alpha_4 -1)\left(\frac{a_1}{a}\right)^{2}
  - \left(\frac{a_1}{a}\right)^{2}  +2 -\alpha_4
 \right]
  \, \mbox{for } a_0 \le a \le a_1 
   \nonumber \\
 f_{\omega\omega}(a) = 
 8   \left(\frac{a_2}{a_1}\right)^3 \left(\frac{a_1}{a}\right)^{2}
\left[(2f_\ell(a_3) - f_\ell(a_1)-27)\alpha_4 - \alpha_4 \left(     -72  \sqrt{\frac{a_1}{a}} +46 \frac{a_1}{a} 
+12  \frac{a_1}{a} \ln{\frac{a}{a_1}}  \right)
\right.
   \nonumber \\
   \left.
+2  \left(  27 -f_\ell(a_3)  -72  \sqrt{\frac{a_1}{a}} +46 \frac{a_1}{a} 
+12  \frac{a_1}{a} \ln{\frac{a}{a_1}}  \right)
 \left(\frac{a_1}{ a}\right)^2\right]
  \, \mbox{for } a_1 \le a \le a_2
   \nonumber \\   
 f_{\omega\omega}(a) = 
 8   \left(\frac{a_2}{a_1}\right)^3  \left(\frac{a_1}{ a}\right)^2
\left\{\left(2f_\ell(a_3) -f_\ell(a_1) -f_\ell(a_2)\frac{a_2}{ a} 
\right.
\right.
   \nonumber \\
   \left.
   -  3 \frac{a_1}{a_2} \left[ 
 \left( 3 \sqrt{\frac{a_2}{a_1}} -\frac{3}{2}\right)^2  \left(1 - \frac{a_2}{a} \right)+\left( 3 \sqrt{\frac{a_2}{a_1}} -\frac{3}{2}\right)  \left(\frac{a_2^2}{a^2} - \frac{a_2}{a} \right)
-\frac{1}{12} \left(\frac{a_2^4}{a^4} - \frac{a_2}{a} \right)
  \right]\right)\alpha_4
   \nonumber \\
   -2\left(f_\ell(a_3)  -f_\ell(a_2)\frac{a_2}{ a}
   \right.
   \nonumber \\
   \left.
   \left.
   -  3 \frac{a_1}{a_2} \left[ 
 \left( 3 \sqrt{\frac{a_2}{a_1}} -\frac{3}{2}\right)^2  \left(1 - \frac{a_2}{a} \right)+\left( 3 \sqrt{\frac{a_2}{a_1}} -\frac{3}{2}\right)  \left(\frac{a_2^2}{a^2} - \frac{a_2}{a} \right)
-\frac{1}{12} \left(\frac{a_2^4}{a^4} - \frac{a_2}{a} \right)
  \right]\right) \left(\frac{a_1}{ a}\right)^2 \right\}
 \nonumber \\  
  \, \mbox{for } a_2 \le a \le a_3
\mbox{ . } \,\,\,\,\,\,\,\,\,\,
\label{jan16.H.f1.sec.order.2}
\end{eqnarray}

If the vorticity is small enough, then we can find a coordinate system with a cosmological scale factor $a$ normal to surfaces of constant global time, that differ from flow lines by a very small amount.  In that case, we can replace the scale factor $\ell$ along flow lines by the cosmological scale factor $a$, and we calculate the effect of that difference to first and second order.  In making this calculation, we replace $\tau$ by the global time $t$.  

We start by using $\dot{\ell} \approx  \dot{a} / \cos{\theta}$  in (\ref{jan16.Friedmann.1.ell}) and neglect shear and acceleration to give
\begin{equation}
\frac{\dot{a}}{a}
 =\frac{\ell}{a}  
  \cos{\theta} 
   \sqrt{ H^2 
 + H_{\omega}^2 + H_{\sigma}^2 + H_{a}^2
  }
\mbox{  }
\label{jan16.Friedmann.1a}
\end{equation}
for the generalized Friedmann equation,
where
\begin{equation}
H \equiv \sqrt{\frac{\Lambda}{3}  + \frac{8\pi \rho}{3}   
 - \frac{k}{{a}^2}   }
 \to
\sqrt{\frac{\Lambda}{3}  + \frac{8\pi \rho}{3}       }
 \mbox{  }
\label{jan16.Hubble-parameter}
\end{equation}
is the Hubble parameter without vorticity, shear, or acceleration, and we neglect the spatial curvature term from here on because measurements show it to be nearly zero.

However, because the density $\rho$ in (\ref{jan16.Hubble-parameter}) depends on $\ell$ rather than $a$, we have in analogy with (\ref{jan16.L.sec.order}) for the same effect on the Lagrangian
\begin{equation}
H^2 \approx H_0^2  \left[1+ f_L(a) \delta + f_{LL}(a) \delta^2 \right]
\mbox{ , }
\label{jan16.H2.sec.order}
\end{equation}
where 
\begin{equation}
H_0 =
\sqrt{\frac{\Lambda}{3}  + \frac{8\pi \rho_0}{3}       }
 \mbox{  }
\label{jan16.Hubble-parameter.0}
\end{equation}
is the Hubble parameter without the $\ell$ dependence and $\rho_0$ is the density without the $\ell$ dependence.  
We also have
\begin{equation}
\frac{1}{H^2} \approx \frac{1}{H_0^2}  \left[1 - f_L(a) \delta + \left(f_L(a)^2-f_{LL}(a)\right) \delta^2 \right]
\mbox{  }
\label{jan16.Hinv2.sec.order}
\end{equation}
and
\begin{equation}
\frac{1}{H} \approx \frac{1}{H_0}  \left[1 - \frac{1}{2}f_L(a) \delta + \left(\frac{3}{8}f_L(a)^2-\frac{1}{2}f_{LL}(a)\right) \delta^2 \right]
\mbox{ . }
\label{jan16.Hinv.sec.order}
\end{equation}

Although in general, the vorticity propagation equation and the shear propagation equation are coupled \cite{Ellis:Elst:1999}, the coupling terms are second-order, so that for very small vorticity and very small shear, we can neglect the coupling.  Therefore, we can neglect the shear term in (\ref{jan16.Friedmann.1a}).  In addition, we neglect the acceleration term.  Using (\ref{jan16.H2.sec.order}) for $H$, (\ref{jan16.H.omega.2.sec.order}) for $H_\omega^2$, (\ref{jan16.ell.6.second.order}) for $\ell$,  and (\ref{jan16.cos.theta}) for $\cos{\theta}$ gives
\begin{eqnarray}
&&
\frac{\dot{a}}{a}
 \approx  
 \nonumber \\ &&
  H_0 \left[1 +\frac{1}{2}\left(f_L(a)+\frac{H_3^2}{H_0^2}f_\omega(a)\right) \delta 
 +\frac{1}{2}\left(f_{LL}(a)+\frac{H_3^2}{H_0^2}f_{\omega\omega}(a)-\frac{1}{4}\left(f_L(a)+\frac{H_3^2}{H_0^2}f_\omega(a)\right)^2\right) \delta^2
 \right] \times
 \nonumber \\ &&
 \left[ 1 + f_\ell (a) \delta+ f_{\ell\ell} (a) \delta^2 \right] \left[ 1-f_\theta(a) \delta +\left(\frac{1}{6}f_\theta^2(a)+\sqrt{f_\theta(a)}f_{\theta\theta}
(a)\right)\delta^2 \right]
\mbox{ . }
\label{jan16.Friedmann.1b}
\end{eqnarray}
Multiplying out and keeping terms through second order in $\delta$ gives
\begin{eqnarray}
&&
\frac{\dot{a}}{a}
 \approx  
  H_0 \left\{1 +\left[\frac{1}{2}f_L(a)+ f_\ell (a)-f_\theta(a)+\frac{1}{2}\frac{H_3^2}{H_0^2}f_\omega(a)\right] \delta 
  \right.
 \nonumber \\ &&
 +\left[\frac{1}{2}f_{LL}(a)+ f_{\ell\ell} (a)+\frac{1}{6}f_\theta^2(a)+\sqrt{f_\theta(a)}f_{\theta\theta} (a)-\frac{1}{8}f_L(a)^2+\frac{1}{2}f_L(a)f_\ell (a)
 -\frac{1}{2}f_L(a)f_\theta(a) -f_\ell (a)f_\theta(a)
   \right.  \nonumber \\ &&  \left.  \left.
 +\left(-\frac{1}{4}f_L(a)f_\omega(a)+\frac{1}{2}f_\ell (a)f_\omega(a)-\frac{1}{2}f_\theta(a)f_\omega(a) +\frac{1}{2}f_{\omega\omega}(a)
 -\frac{1}{8}\frac{H_3^2}{H_0^2}f_\omega(a)^2
 \right) \frac{H_3^2}{H_0^2}
 \right] \delta^2
 \right\} 
 \mbox{ . }
\label{jan16.Friedmann.1d}
\end{eqnarray}

To the same degree of approximation, we have
\begin{eqnarray}
&&
\frac{1}{\dot{a}}
 \approx  
 \nonumber \\ &&
  \frac{1}{a H_0} \left[1 -\frac{1}{2}\left(f_L(a)+\frac{H_3^2}{H_0^2}f_\omega(a)\right) \delta 
 -\frac{1}{2}\left(f_{LL}(a)+\frac{H_3^2}{H_0^2}f_{\omega\omega}(a)-\frac{3}{4}\left(f_L(a)+\frac{H_3^2}{H_0^2}f_\omega(a)\right)^2\right) \delta^2
 \right] \times
 \nonumber \\ &&
 \left[ 1 - f_\ell (a) \delta+ \left(f_\ell^2 (a)-f_{\ell\ell} (a)\right) \delta^2 \right] 
 \left[ 1+f_\theta(a) \delta +\left(\frac{5}{6}f_\theta^2(a)-\sqrt{f_\theta(a)}f_{\theta\theta}
(a)\right)\delta^2 \right]
\mbox{ . }
\label{jan16.Friedmann.1b.inv}
\end{eqnarray}
Multiplying out and keeping terms through second order in $\delta$ gives
\begin{equation}
\frac{1}{\dot{a}}  \approx    \frac{1}{a H_0}  \left[1+f_H(a) \delta +f_{HH}(a) \delta^2\right]
\mbox{ , }
\label{jan16.Friedmann.1d.inv.2}
\end{equation}
where
\begin{equation}
f_H(a) = -\frac{1}{2}f_L(a)- f_\ell (a)+f_\theta(a)-\frac{1}{2}\frac{H_3^2}{H_0^2}f_\omega(a)
\mbox{  }
\label{jan16.Friedmann.f.H}
\end{equation}
and
\begin{eqnarray}
&&
f_{HH}(a) = 
-\frac{1}{2}f_{LL}(a)
+ f_{\ell}^2 (a)
- f_{\ell\ell} (a)
+\frac{5}{6}f_\theta^2(a)
-\sqrt{f_\theta(a)}f_{\theta\theta} (a)
+\frac{3}{8}f_L(a)^2
\nonumber \\ &&
+\frac{1}{2}f_L(a)f_\ell (a)
 -\frac{1}{2}f_L(a)f_\theta(a)
  -f_\ell (a)f_\theta(a)
   \nonumber \\ &&  
 +\left(\frac{3}{4}f_L(a)f_\omega(a)+\frac{1}{2}f_\ell (a)f_\omega(a)-\frac{1}{2}f_\theta(a)f_\omega(a) -\frac{1}{2}f_{\omega\omega}(a)
 +\frac{3}{8}\frac{H_3^2}{H_0^2}f_\omega(a)^2
 \right) \frac{H_3^2}{H_0^2}
\mbox{ . }
\label{jan16.Friedmann.f.HH}
\end{eqnarray}

We can consider that (\ref{jan16.Friedmann.1d}) gives the effective generalized Friedmann equation, including to second order the effect that flow lines are not normal to surfaces of constant global time.